\documentclass[reprint,showpacs,superscriptaddress,aps,twocolumn,epsfig,showkeys]{revtex4-2}
\usepackage{amsmath,amssymb,graphicx}
\usepackage{latexsym}
\usepackage{esint}
\usepackage{color}
\usepackage{float}
\usepackage{multirow}
\usepackage{hyperref}
\usepackage[normalem]{ulem}
\usepackage{enumerate}
\usepackage{amsbsy}
\usepackage{bm}
\usepackage{ulem}

\newcommand{\be}{\begin{equation}}
\newcommand{\ee}{\end{equation}}
\newcommand{\bse}{\begin{subequations}}
\newcommand{\ese}{\end{subequations}}
\newcommand{\bea}{\begin{eqnarray}}
\newcommand{\eea}{\end{eqnarray}}
\newcommand{\ba}{\begin{array}}
\newcommand{\ea}{\end{array}}

\usepackage[usenames,dvipsnames]{xcolor}
\hypersetup{
colorlinks,
citecolor=blue,
filecolor=blue,
linkcolor=blue,
urlcolor=blue}

\newenvironment{packed_enum}{
\begin{itemize}
  \setlength{\itemsep}{0pt}
  \setlength{\parskip}{0pt}
  \setlength{\parsep}{0pt}
}{\end{itemize}}

\usepackage{nicefrac}

\begin{document}

\title{The geometric phase transition of the three-dimensional $\mathbb{Z}_2$ lattice gauge model}

\author{Ramgopal Agrawal,$^1$ Leticia F. Cugliandolo,$^{1}$ Lara Faoro,$^2$ Lev B. Ioffe,$^2$ and Marco Picco$^1$\\
	{\small $^1$Sorbonne Universit\'e, Laboratoire de Physique Th\'eorique et Hautes Energies}, \\
	{\small CNRS UMR 7589, 4 Place Jussieu, 75252 Paris Cedex 05, France}
	\\
	{\small $^2$Google Research, Mountain View, California 94043, USA}
}

\date{\today}

\begin{abstract}
After fifty years of \textit{lattice gauge theories} (LGTs), the nature of the transition between their topological phases (confinement/deconfinement) remains challenging due to the absence of a local order parameter. In this work, we conduct a percolation analysis of Wegner's three-dimensional $\mathbb{Z}_2$ lattice gauge model using intensive Monte Carlo simulations and finite-size scaling, offering fresh insights into the topological phase transitions of gauge-invariant systems. We demonstrate that, regardless of the connection rules, geometrical loops, constructed by piercing excited plaquettes percolate precisely at the thermal critical point $T_{\rm c}$, with critical exponents coinciding with those of the loop representation of the dual 3D Ising model. Further, we construct Fortuin-Kasteleyn (FK) clusters in a random-cluster representation, showing that they also percolate at $T_{\rm c}$, enabling access to all thermal critical exponents. Strikingly, the Binder cumulants of the percolation order parameters for both loops and FK clusters reveal a \textit{pseudo-first-order} transition. This work sheds new light on the critical behavior of pure LGTs, with potential implications for condensed matter systems and quantum error correction.
\end{abstract}

\maketitle

\textit{Lattice gauge theories} (LGTs) were proposed by Wilson in 1974~\cite{PhysRevD.10.2445} as a simplification of gauge field theories, in which the Euclidean spacetime is discretized on a lattice while preserving the local gauge invariance. The gauge fields represented by a dynamical variable are placed on the links of the lattice, whereas the matter fields are located on its sites. This allowed for the exploration of, e.g., QCD and Yang-Mills models using Monte Carlo sampling methods~\cite{RevModPhys.55.775} and intriguing applications to condensed matter theory~\cite{Fradkin,Gregor_2011,Sachdev_2019}.

Interestingly, prior to Wilson’s work, Wegner~\cite{10.1063/1.1665530} had already introduced in 1971 a gauge-invariant version of the classical Ising model (IM). This model, a pure gauge theory with $\mathbb{Z}_2$ degrees of freedom and no matter fields, demonstrated that phase transitions of topological nature can occur without a local order parameter.

In the last two decades, pure gauge lattice models have played a key role in topological codes, providing a framework for studying error correction. The most prominent example is the Toric Code~\cite{Kitaev_1997,KITAEV20032}, where error chain recovery undergoes a geometric phase transition. Accurate estimates of the error threshold arise from mapping to the order-disorder transition of classical disordered spin models along the Nishimori line~\cite{doi:10.1063/1.1499754,nishimori2001statistical}. For Toric Codes with \textit{realistic} errors, the corresponding model is the 3D $\mathbb{Z}_2$ LGT with quenched randomness~\cite{doi:10.1063/1.1499754,WANG200331,OHNO2004462}. These connections have sparked renewed interest in phase transitions in pure LGTs and Toric Codes~\cite{Andrist_2011,PhysRevA.94.012318,PhysRevA.85.050302,PhysRevB.78.155120,PhysRevLett.106.107203,PhysRevLett.112.070501,PhysRevLett.127.235701,PhysRevResearch.3.043209,PRXQuantum.3.030338,apte2024deeplearninglatticegauge}.

Despite its long history, identifying and characterizing phase transitions between topological phases in a \textit{pure} LGT remains challenging. Without a local order parameter, the correlation length exponent $\nu$ can be extracted from global gauge-invariant observables~\cite{Bricmont1980}, but exponents like $\beta$ and $\gamma$ lack clear physical interpretation. Moreover, global quantities obscure precise detection of the transition point in gauge-invariant models.

Recent progress in pure LGTs~\cite{OHNO2004462,Andrist_2011,PhysRevA.94.012318,PhysRevB.97.024432} has relied on specific heat or global order parameters, such as Wilson~\cite{PhysRevD.10.2445,RevModPhys.51.659} and Polyakov loops~\cite{POLYAKOV197582,SVETITSKY1982423}, to probe the transition. Yet neither approach offers a complete description. This underscores the need for new methods to investigate critical phenomena under gauge invariance -- methods also relevant to topological transitions in gauge-invariant condensed matter systems~\cite{PhysRevB.44.2664,doi:10.1126/science.1091806,PhysRevB.72.045141,PhysRevLett.98.106803,Gregor_2011,doi:10.1142/S0217979212300071,Sachdev_2019}.

A natural route is to identify objects whose fractal and statistical properties fully characterize the phase transition. A plethora of works have analyzed geometrically defined objects across diverse 3D systems. Some examples are cosmic strings in the early universe~\cite{Kibble_1976,PhysRevD.30.2036}, lines of darkness in light field~\cite{PhysRevLett.100.053902}, vortex loops in $XY$ system~\cite{PhysRevLett.57.1358,KAJANTIE2000114}, complex $\vert \psi \vert^4$ theories~\cite{PhysRevB.72.094511,PhysRevE.94.062146,Kobayashi_2016}, etc. Although these objects exhibit a unique geometric phase transition, their percolation temperature $T_{\rm p}$ and critical properties do not always agree with the ones of the thermodynamic transition~\cite{MULLERKRUMBHAAR197427}. For instance, the domains formed by parallel nearest neighboring spins in the 3DIM do not percolate at the thermodynamic critical temperature $T_c$, i.e., $T_{\rm p} \ne T_{\rm c}$. Instead, the clusters constructed using the random-cluster representation~\cite{FORTUIN1972536}, known as the \textit{Fortuin-Kasteleyn} (FK) clusters, do percolate at $T_{\rm c}$~\cite{Coniglio_1980,DOTSENKO1995577,PhysRevE.99.042150} and their critical exponents agree with the thermal critical ones.

In this Letter we study the celebrated 3D Wegner plaquette model from a geometric perspective. We analyze loops formed by joining line segments that pierce frustrated plaquettes perpendicularly. These loops in our construction are  the center vortices of the model (see Refs.~\cite{GREENSITE20031,PhysRevD.61.054504,PhysRevD.66.017501,PhysRevD.111.074512} for extensive discussion). Additionally, we introduce a FK formulation involving larger graphs and investigate these in detail. With this study we unveil a full description of the thermal critical properties in which both kinds of objects play a relevant role.

{\it Model and analysis.}
Wegner's 3D $\mathbb{Z}_2$ gauge model is defined by the Hamiltonian
\be
H = -J \sum_{\rm P} U_{\rm P} = -J \sum_{\rm P} \prod_{\ell \in P} S_{\ell}
\; ,
\label{eq1}
\ee
where $S_{\ell}(=\pm 1)$ are Ising spins placed on the links, $\ell$, of the square plaquettes, ${\rm P}$, of the cubic lattice, and $J > 0$ is a coupling constant. The contribution from each plaquette, $U_{\rm P} = \prod_{\ell \in {\rm P}} S_{\ell}$, takes either $+1$ (simple plaquette) or $-1$ (frustrated plaquette) values. Due to local gauge invariance~\footnote{A local gauge transformation flips all link spins at a site, leaving $U_{\rm P}$ and the Hamiltonian unchanged.}, each plaquette configuration (including the ground state $U_{\rm P}=+1 \,  \forall{\rm P}$) is enormously degenerate, a phenomenon also referred to as \textit{gauge redundancy}.

The equilibrium magnetization density of the model~\eqref{eq1} vanishes at all $T/J \ge 0$. Some understanding of the phase transition is gained by Kramers–Wannier duality~\cite{10.1063/1.1665530,PhysRevD.11.2098,PhysRevD.21.2892}. In 3D, the ${\mathbb Z}_2$ gauge model is dual to the ferromagnetic IM. The critical inverse-temperature of the gauge model $\beta_c$ is related to the one of the conventional IM $\beta_c^*$ through $\beta_c J = -(\ln \tanh (\beta^*_c J^*))/2 $. This yields $T_c \simeq 1.313346$ $J/k_{\rm B}$. The duality also corroborates~\cite{PhysRevD.11.2098,PhysRevD.21.2892,KEHL1988324,CASELLE1996435} that the thermal critical exponents should be the same as those in the dual model. Without a local order parameter, nonlocal gauge-invariant quantities characterize the two phases: low-temperature \textit{deconfinement} and high-temperature \textit{confinement}; see Supplemental Material (SM).

After defining the two geometric objects under study, we apply a detailed finite size scaling analysis following percolation studies~\cite{stauffer2018introduction,Essam_1980,PhysRevLett.53.1121,SABERI20151,Strelniker2009}. With it, we
\begin{packed_enum}
	\item[(i)] identify $T_{\rm p}$ and $1/\nu_{\rm p}$ from the spanning probability $P_{\rm S}$ (fraction of configurations with at least one spanning object) and the Binder cumulant $U_{\rm 4}$ of the mass fraction $m$ of the largest object.
	\item[(ii)] obtain $\beta_{\rm p}/\nu_{\rm p}$ from the average mass fraction of the largest object, or percolation strength, $P$.
	\item[(iii)] deduce $\gamma_{\rm p}/\nu_{\rm p}$ from the susceptibility $\chi$ defined via fluctuations in $m$.
	\item[(iv)] confirm the robustness of analysis via cluster size distributions (yielding $\tau_{\rm p}$ and $\sigma_{\rm p}$) and different measures of fractal dimension $D_{\rm f}$.
\end{packed_enum}
The precise definitions of all these geometric observables are given in the SM.

\begin{figure}[t!]
	\centering
	\rotatebox{0}{\resizebox{.49\textwidth}{!}{\includegraphics{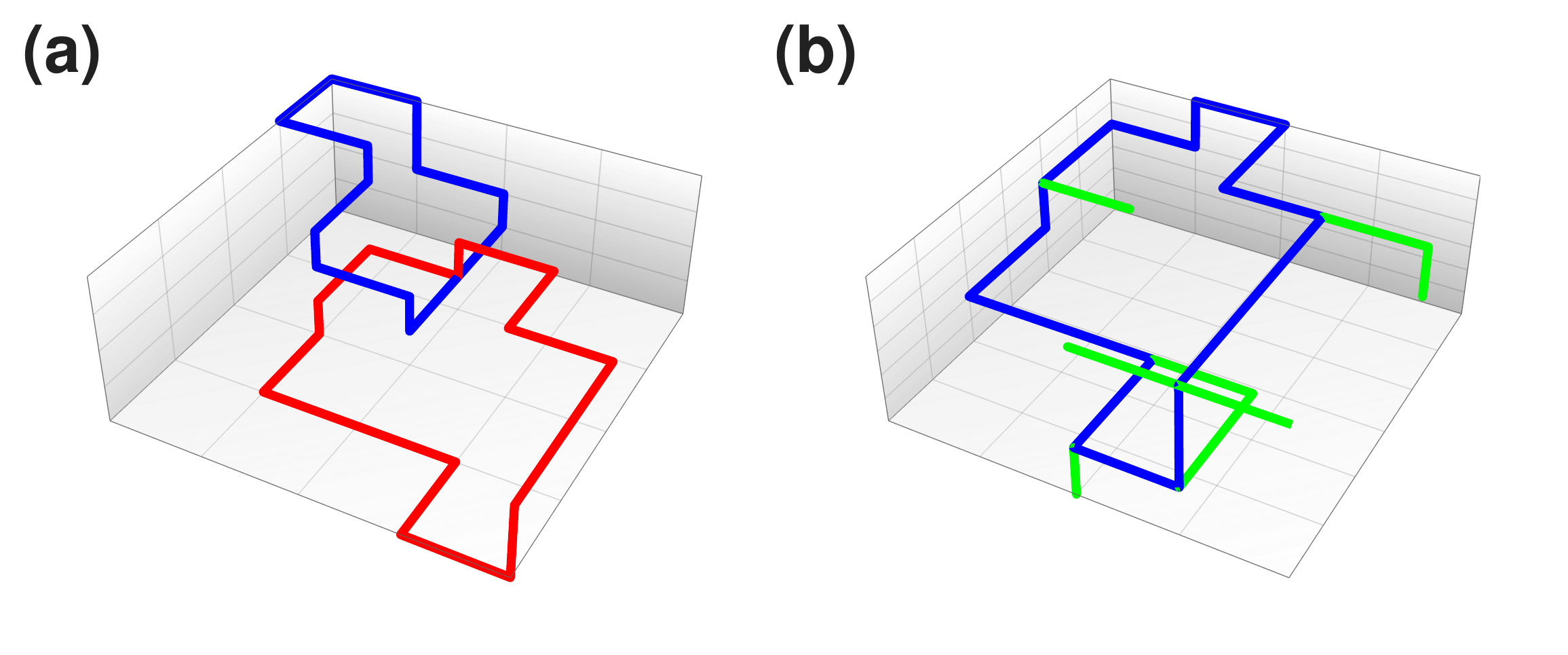}}}
	\caption{Schematic of (a) line loops and (b) FK graphs in the 3D gauge model. In (a), colors indicate distinct loops piercing frustrated plaquettes. In (b), the graph is formed by piercing empty plaquettes, with dark blue segments for ``frustrated" plaquettes and light green segments for ``simple" ones added to the loop.}
	\label{fig1}
\end{figure}

\begin{figure*}[t!]
	\centering
	\rotatebox{0}{\resizebox{.99\textwidth}{!}{\includegraphics{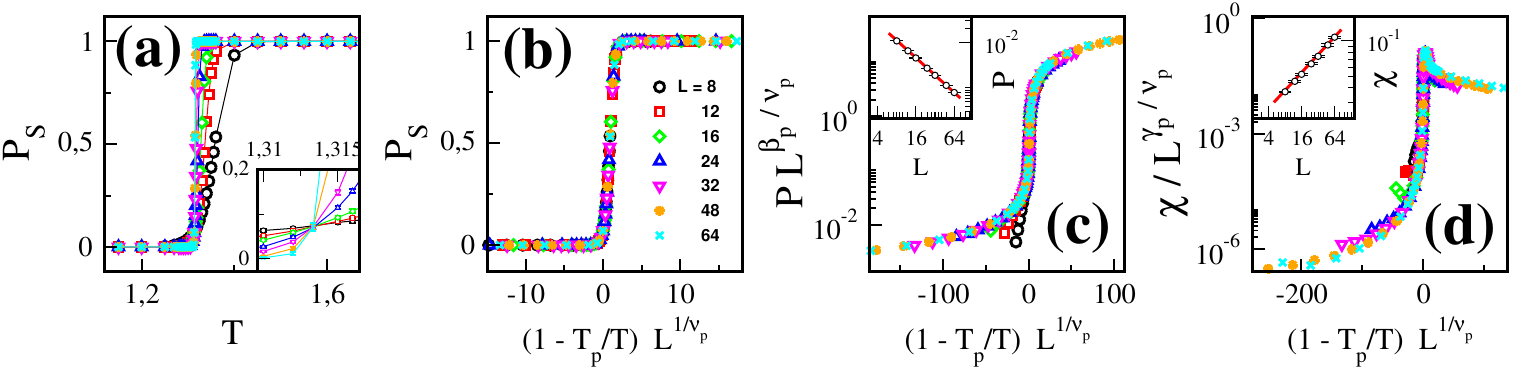}}}
	\caption{{\bf Geometric line loops}. Spanning probability $P_{\rm S}$ vs temperature $T$ (a) and 
		against the scaling variable $(1-T_{\rm p}/T) L^{1/\nu_{\rm p}}$ (b). (c) Rescaled plot of the percolation strength, $P L^{\beta_{\rm p}/\nu_{\rm p}}$ vs $(1-T_{\rm p}/T) L^{1/\nu_{\rm p}}$. (d) Scaling plot of the susceptibility, $\chi / L^{\gamma_{\rm p}/\nu_{\rm p}}$ vs $(1-T_{\rm p}/T) L^{1/\nu_{\rm p}}$. Different datasets represent systems of different linear size $L$, see the key in (b). The inset in (a) magnifies the intersection region in the main frame. The insets in (c) and (d) display the values of $P$ and $\chi$ at $T_{\rm c}$ against the system size $L$, with solid lines representing the laws $P \sim L^{-\beta_{\rm p}/\nu_{\rm p}}$ and $\chi \sim L^{\gamma_{\rm p}/\nu_{\rm p}}$, respectively. In (b)-(d),  $T_{\rm p} = 1.3133$, $1/\nu_{\rm p} = 1.58$, $\beta_{\rm p}/\nu_{\rm p} = 1.26$, and $\gamma_{\rm p}/\nu_{\rm p} = 0.47$. 
		The collapses hold within the numerical error bars; see Table~\ref{tab1}.}
	\label{fig2}
\end{figure*}

{\it Geometric line loops.}
The plaquettes with $U_{\rm P} = -1$ are excitations on top of the ground state which are generated by thermal fluctuations. In a dual representation each frustrated plaquette is pierced by a perpendicular segment with lattice spacing length. Since each link spin is shared by two faces of the same cube, only an even number of plaquettes on the faces of an elementary cube can have $U_{\rm P} = -1$~\footnote{This follows from the discrete Bianchi identity, which forbids monopoles.}. These segments are then joined to form closed line loops. When more than two plaquettes within a cube are frustrated, e.g., four or six, that cube forms a branch point, introducing ambiguity in loop identification; see the schematic in Fig.~\ref{fig1}(a). At low temperatures the loops are short while at higher temperatures, due to stronger thermal agitation, much longer and even giant system-sized loops can be formed. This suggests a transition from finite loops to system-spanning loops~\cite{PhysRevD.66.017501,PhysRevLett.112.070501,PhysRevD.111.074512}, underlying the need of a careful percolation analysis. As explained in the SM, these loops correspond to graphs in the low-temperature (LT) expansion of the partition function and, by duality, match the high-temperature (HT) expansion graphs of the 3DIM.

\begin{figure}[b!]
	\centering
	\rotatebox{0}{\resizebox{.48\textwidth}{!}{\includegraphics{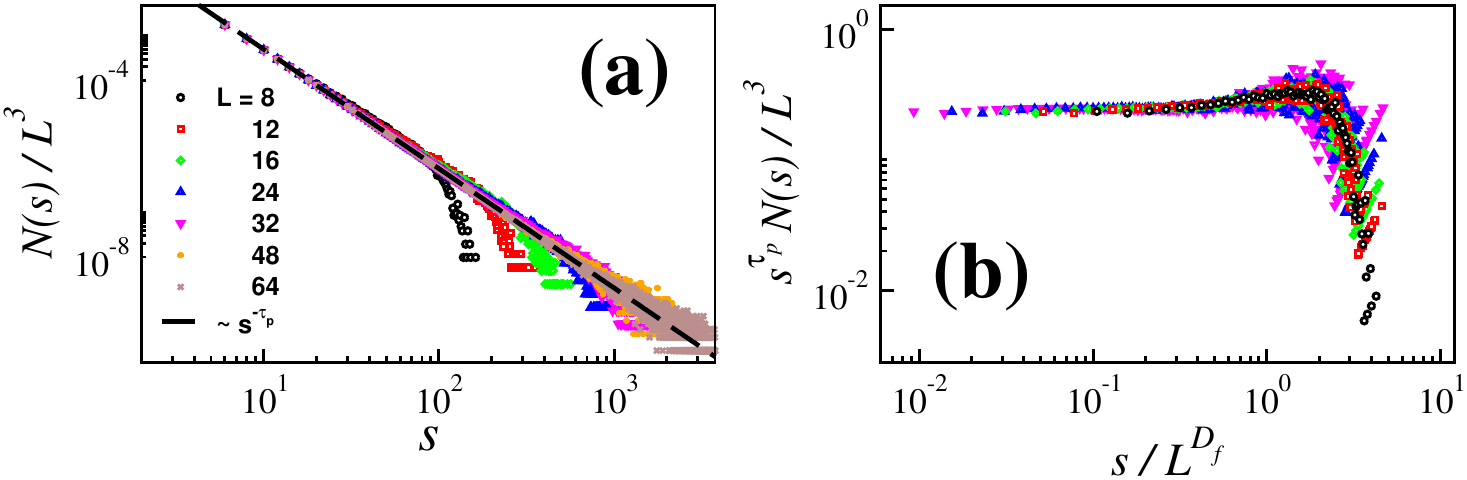}}}
	\caption{{\bf Geometric line loops}. (a) Number density $N(s)/L^3$ vs mass $s$ on a double logarithmic scale for different $L$ (see the key) at $T_{\rm c}$. The dashed line represents the law $s^{-\tau_{\rm p}}$, with $\tau_{\rm p} = 2.71$. (b) plots the scaling variable $s^{\tau_{\rm p}} N(s)/L^3$ against $s/L^{D_{\rm f}}$ for data in (a), wherein $\tau_{\rm p}$ and $D_{\rm f}$ are fixed to $\tau_{\rm p} = 2.71$ and $D_{\rm f} = 1.74$.}
	\label{fig3}
\end{figure}

Before proceeding, we discuss how to unambiguously identify closed loops at branch points. In the literature, two methods are proposed~\cite{KAJANTIE2000114,PhysRevB.72.094511,PhysRevE.94.062146} to resolve the branches. Both forbid crossings between different loops. One of the methods is the ``maximal reconnection method'' in which the connections at a branching point are chosen so as to maximize loop length. The other one is the ``stochastic method'' in which a dice is used to stochastically determine the connections. In our system, both methods give equivalent results~\footnote{In the stochastic method, loops are generally smaller than those from the maximal rule. At $T \le T_{\rm c}$, the scarcity of branch points leads to similar loop statistics across methods. In contrast, for $T>T_{\rm c}$, notable differences emerge; see SM and Refs.~\cite{de1979scaling,PhysRevLett.107.177202,PhysRevLett.111.100601}.}. Here we show data for the maximal reconnection rule only. In the SM we complement the analysis with data built with the stochastic rule. Further technical details of the Monte Carlo (MC) simulations~\cite{newman1999monte} and the properties of other observables of interest are also relegated to the SM.

Figure~\ref{fig2} presents the analysis of geometric line loops built using the maximal connectivity rule. In (a), the spanning probability $P_{\rm S}$ is plotted against temperature for various system sizes $L$. At low $T$, $P_{\rm S} \to 0$, while at high $T$, $P_{\rm S} \to 1$. At $T_{\rm p}$, $\lim_{L \to \infty} P_{\rm S}$ reaches a universal value~\cite{langlands1992universality,PhysRevLett.69.2670,PhysRevE.53.235}. The intersection of curves gives $T_{\rm p} \simeq 1.3134(3)$ (inset), in good agreement with $T_{\rm c}$ from duality. A similar match is found using the stochastic rule [see SM]. For a more stringent test, Fig.~\ref{fig2}(b) plots $P_{\rm S}$ versus the scaling variable $(1 - T_{\rm p}/T) L^{1/\nu_{\rm p}}$. Fixing $T_{\rm p} = T_{\rm c}$ and fitting $\nu_{\rm p}$ yields excellent scaling collapse~\cite{*[{We use a local-linearity function, akin to reduced $\chi^2$, to quantify collapse quality. See~}][{}] doi:10.1143/JPSJ.62.435} with $1/\nu_{\rm p} \simeq 1.59(3)$, consistent with the 3D Ising class~\cite{PhysRevB.82.174433,kos2016precision,simmons2017lightcone,PhysRevE.97.043301}. Analysis of the Binder cumulant $U_{\rm 4}$ of the largest object's mass fraction $m$ also supports these values; see the SM and discussion below.

\begin{figure*}[t!]
	\centering
	\rotatebox{0}{\resizebox{.95\textwidth}{!}{\includegraphics{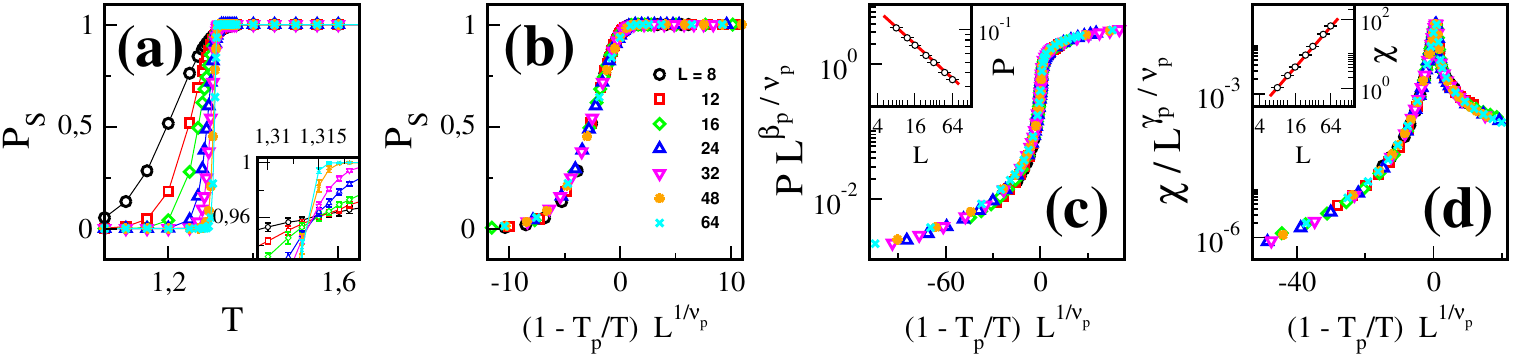}}}
	\caption{{\bf Fortuin-Kasteleyn graphs}. Various panels show plots similar to the ones in Fig.~\ref{fig2}. In (b)-(d),
		$T_{\rm p} = 1.3133$, $1/\nu_{\rm p} = 1.58$, $\beta_{\rm p}/\nu_{\rm p} = 0.52$, and $\gamma_{\rm p}/\nu_{\rm p} = 1.97$, 
		and the collapses hold within the numerical error bars; see Table~\ref{tab1}.}
	\label{fig4}
\end{figure*}

Panel (c) of Fig.~\ref{fig2} displays the scaling plot of the percolation strength $P$, which serves as the order parameter of the percolation transition. The ratio $\beta_{\rm p}/\nu_{\rm p}$ was extracted from the decay $P \sim L^{-\beta_{\rm p}/\nu_{\rm p}}$ at the critical temperature identified in (a)-(b). The excellent collapse in the main panel, using this value, confirms the coherence of the procedure. A similar analysis is shown in (d) for the susceptibility $\chi$: we obtained $\gamma_{\rm p}/\nu_{\rm p}$ from $\chi \sim L^{\gamma_{\rm p}/\nu_{\rm p}}$ at $T_c$, with results consistent with the scaling collapse. The fractal dimension $D_{\rm f}$ was determined from the average mass $\langle s_{l} \rangle$ of the largest object. The values of $T_{\rm p}$, $1/\nu_{\rm p}$, $\beta_{\rm p}/\nu_{\rm p}$, $\gamma_{\rm p}/\nu_{\rm p}$, and $D_{\rm f}$ are listed in Table~\ref{tab1} and match those of HT graphs in the 3DIM within numerical precision~\cite{PhysRevE.77.061108,shimada2016fractal,PhysRevE.101.012104}. Notably, for the loop representation, $1/\nu_{\rm p}$ is associated with the scaling dimension $\Delta_{\epsilon}$ of the energy operator $\epsilon$ in the 3D ${\cal O}(N=1)$ singlet sector, while $\beta_{\rm p}/\nu_{\rm p}$ and $\gamma_{\rm p}/\nu_{\rm p}$ correspond to the scaling dimension $\Delta_T$ of a \textit{nontrivial} symmetric tensor operator $\varphi_{ab}$ present only in the 3D ${\cal O}(N \to 1^+)$ tensor sector~\cite{shimada2016fractal,PhysRevE.101.012104}; see the SM.

\begin{table}[b!]
	\begin{center}
		\begin{tabular}{ | c || c | c | c | c | c |}
			\hline
			\hline
			\footnotesize{System} & \ $T_{\rm p}$ or $T_{\rm c}$ \ & \ \ $\nicefrac{1}{\nu_{\rm p}}$ \ \ & \ \ $\nicefrac{\beta_{\rm p}}{\nu_{\rm p}}$ \ \ & \ \ $\nicefrac{\gamma_{\rm p}}{\nu_{\rm p}}$ \ \ & \ \ $D_{\rm f}$ \ \ \\
			\hline
			\hline
			\scriptsize{FK IM\cite{PhysRevB.82.174433,kos2016precision,simmons2017lightcone,PhysRevE.97.043301}} & 1.3133 & 1.587 & 0.518 & 1.964 & 2.482 \\
			
			\scriptsize{HT IM\cite{PhysRevE.77.061108,shimada2016fractal,PhysRevE.101.012104}} & 1.3133 & 1.587 & 1.265 & 0.470 & 1.735 \\
			
			\footnotesize{Loops GM} & 1.3134(3) & 1.592(28) & 1.261(8) & 0.486(18) & 1.738(8) \\
			
			\footnotesize{FK GM} & 1.3137(4) & 1.581(18) & 0.519(5) & 1.971(7) & 2.481(6) \\
			\hline
			\hline
		\end{tabular}
	\end{center}
	\caption{Comparison of FK clusters in 3DIM (FK IM), HT expansion graphs in 3DIM (HT IM), geometric loops (Loops GM), and FK clusters (FK GM) in 3D gauge model. Estimates for the percolation temperature $T_{\rm p}$, exponents $1/\nu_{\rm p}$, $\beta_{\rm p}/\nu_{\rm p}$, $\gamma_{\rm p}/\nu_{\rm p}$, and fractal dimension $D_{\rm f}$. For FK IM and HT IM, the value of $T_{\rm c} (\equiv 1/\beta_c)$ is given instead of its dual.}
	\label{tab1}
\end{table}

The number density of loops with mass $s$ is displayed in Fig.~\ref{fig3}. The algebraic decay $N(s)/L^3 \sim s^{-\tau_{\rm p}}$ is evident from (a), with a fit yielding the Fisher exponent $\tau_{\rm p} \sim 2.71(1)$. Using the relation $\tau_{\rm p} = 1 + D/D_{\rm f}$, we obtain $D_{\rm f} = 1.74(1)$, consistent with value from $\langle s_{l} \rangle$ despite larger error bars. Panel (b) confirms scaling collapse in agreement with these estimates.

We remark that within numerical accuracy the relations $\beta_{\rm p}/\nu_{\rm p} = D-D_{\rm f}$ and $\gamma_{\rm p}/\nu_{\rm p} = 2D_{\rm f} - D$ are also obeyed, providing an independent check for our estimates. In conclusion, the line loops are equivalent to the HT expansion graphs of the 3DIM.

{\it Fortuin-Kasteleyn (FK) graphs.}
Likewise the Ising model, we construct the FK clusters in model~\eqref{eq1} by using the random cluster representation~\cite{FORTUIN1972536,Coniglio_1980}. Herein, a bimodal random variable ($n_{\rm P}=0,1$) is assigned to each plaquette on the lattice such that a plaquette is occupied if $n_{\rm P} = 1$ and it is empty if $n_{\rm P} = 0$; see SM for details. The partition function is given by
\bea
{\cal Z} 
&=& \!\! \sum_{\text{config.}} \prod_{\rm P} \sum_{n_{\rm P} = 0,1} \! {\rm e}^{\beta J} \left[ (1-q) \delta_{n_{\rm P},0} + q \delta_{U_{\rm P},1} \delta_{n_{\rm P},1} \right]
\, ,
\;\;\;\;
\eea
where $q \equiv 1 - {\rm e}^{-2\beta J}$. A plaquette with $U_{\rm P} = -1$ is always empty in this representation, while the one with $U_{\rm P} = 1$ is empty with probability $(1-q)$. Thus, a FK cluster is formed by the set of empty plaquettes which are connected to each other.  The linear structures piercing the FK clusters form closed loops with attached linear protrusions; see the Schematic Fig.~\ref{fig1}(b). These graphs will be typically larger than the geometric loops and are not completely closed.

In Fig.~\ref{fig4} we apply the percolation analysis on the FK graphs. The finite-size scaling quality is again very good. The values of $T_{\rm p}$ and $1/\nu_{\rm p}$ match, within numerical accuracy, those found for geometric loops. However, the values of $\beta_{\rm p}/\nu_{\rm p}$ and $\gamma_{\rm p}/\nu_{\rm p}$ in Table~\ref{tab1} differ markedly from the geometric case and coincide with those of FK clusters in the 3DIM. This is because these FK exponents in the gauge model are related to the scaling dimension $\Delta_{\sigma}$ of the \textit{singlet} spin operator $\sigma$ of the 3DIM~\cite{shimada2016fractal}.

The analysis of the number density and fractal properties of the FK graphs, analogous to Fig.~\ref{fig3} for loops, yields the value of $D_{\rm f}$ reported in Table~\ref{tab1}. It also satisfies the scaling relations; see the SM.

\begin{figure}[b!]
	\centering
	\rotatebox{0}{\resizebox{0.49\textwidth}{!}{\includegraphics{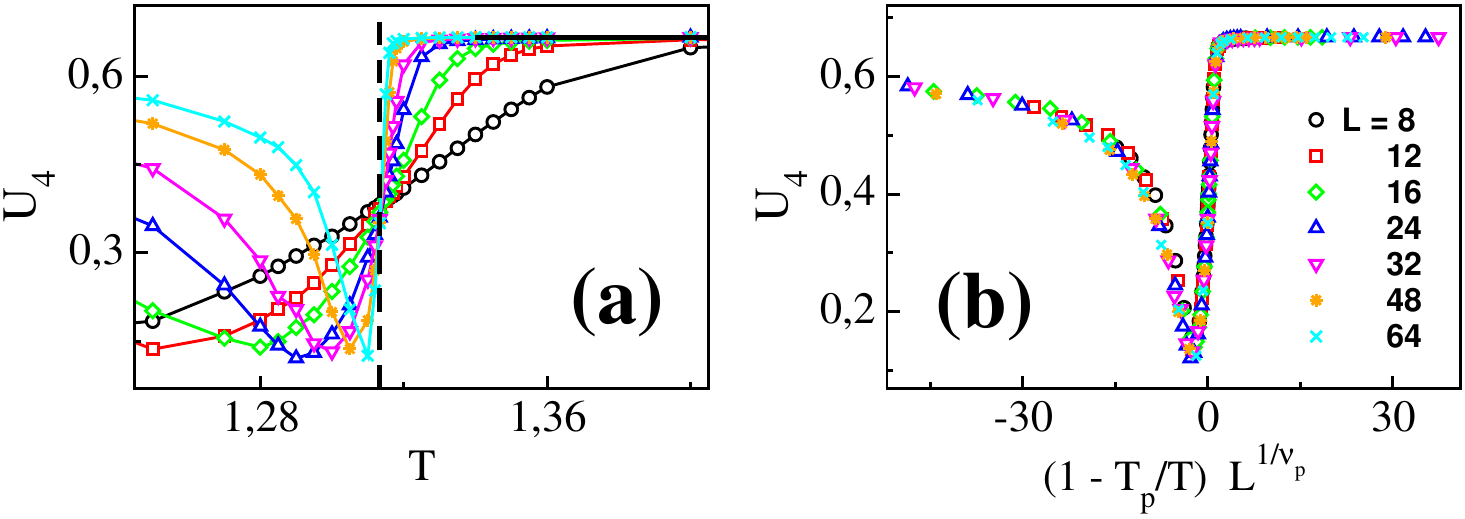}}}
	\caption{{\bf Binder cumulant} $U_{\rm 4}$ for the largest FK cluster. (a) Zoom around $T_{\rm c}$, signaled with a vertical dashed line. The horizontal solid line represents the large-$T$ behavior, $U_{\rm 4} = 2/3$. (b) Scaling of the data fixing $T_{\rm p} = 1.3133$ and $1/\nu_{\rm p} = 1.58$. The linear sizes $L$ are given in the key.
	}
	\label{fig5}
\end{figure}

Finally, we discuss the nature of the geometric transition. The scaling analysis shows that the percolation transition in both geometric loops and FK graphs is continuous. However, the Binder cumulant $U_{\rm 4}$ of $m=s_l/(3L^3)$ has a peculiar behavior. For both the loops and FK graphs, right below $T_{\rm c}$, a dip is observed in $U_{\rm 4}$, see Fig.~\ref{fig5}(a) for the FK case, which does not vary much with system size. If this dip were to increase rapidly with system size, the transition would likely be first order~\cite{PhysRevLett.47.693,PhysRevB.30.1477}. Moreover, the data satisfy finite size scaling $U_{\rm 4} = f_U[(1-T_{\rm p}/T) L^{1/\nu_{\rm p}}]$ (Fig.~\ref{fig5}(b)), as in a continuous transitions. The dip in $U_{\rm 4}$ is due to anomalies in the order parameter distribution $P_{\text{dist}}(m)$ below $T_{\rm c}$ (see SM), characteristic of \textit{pseudo-first-order} transitions~\cite{PhysRevLett.108.045702,PhysRevX.7.031052}.

{\it Conclusion.}
We studied the geometric properties of closed loops formed by frustrated plaquettes in Wegner's 3D $\mathbb{Z}_2$ lattice gauge model. Using different connection rules, we confirmed that these loops percolate at the critical temperature $T_c$, regardless of the specific rule used. Finite-size scaling of percolation observables yields critical exponents matching those of HT graphs of the dual 3DIM. We also identified FK clusters via the random-cluster representation and showed that they also percolate at $T_c$, with exponents matching the thermal ones of the dual 3DIM. Both objects probe scaling dimensions of various operators within a common 3D ${\cal O}(N\to 1^+)$ CFT~\cite{shimada2016fractal,PhysRevE.101.012104}, indicating that they belong to the same universality class. Interestingly, the non-Gaussian behavior of the order parameter distributions below $T_c$ differs from that of conventional second-order transitions, and is characteristic of a pseudo-first-order transition.

This analysis introduces a novel quantitative approach to deconfinement-confinement transitions in pure LGTs, offering new insights into their continuous topological nature and, to our knowledge, the first direct calculation of various critical exponents in this model.

The connection between closed loops and center vortices~\cite{PhysRevD.66.017501,PhysRevD.111.074512}, as well as defect clusters in the Toric code~\cite{PhysRevLett.112.070501}, highlights the broader significance of this work. Moreover, our findings also clarify the relevance of FK clusters for MC methods, resolving earlier concerns about detailed balance and criticality~\cite{ben1990critical}, and supporting their use to reduce critical slowing down.

\begin{acknowledgments}
The authors acknowledge financial support from the French grant ANR-19-CE30-0014 and a Google Gift. The simulations were performed on the SACADO MeSU platform at Sorbonne Universit\'e. RA thanks E. Trevisani and K. J. Wiese for useful discussions. We also thank M. Caselle for stimulating conversation.
\end{acknowledgments}

\bibliography{ref.bib}

\newpage
\section*{Supplemental Material}

\setcounter{section}{0} % Reset section counter
\renewcommand\thesection{S\arabic{section}}

\setcounter{figure}{0} % Reset figure counter
\renewcommand\thefigure{S\arabic{figure}}
\renewcommand\theHfigure{S\arabic{figure}}

\setcounter{equation}{0} % Reset equation counter
\renewcommand\theequation{S\arabic{equation}}

\section{Technical details and methodology}
\label{ssec1}

\begin{figure}[t!]
	\centering
	\rotatebox{0}{\resizebox{.45\textwidth}{!}{\includegraphics{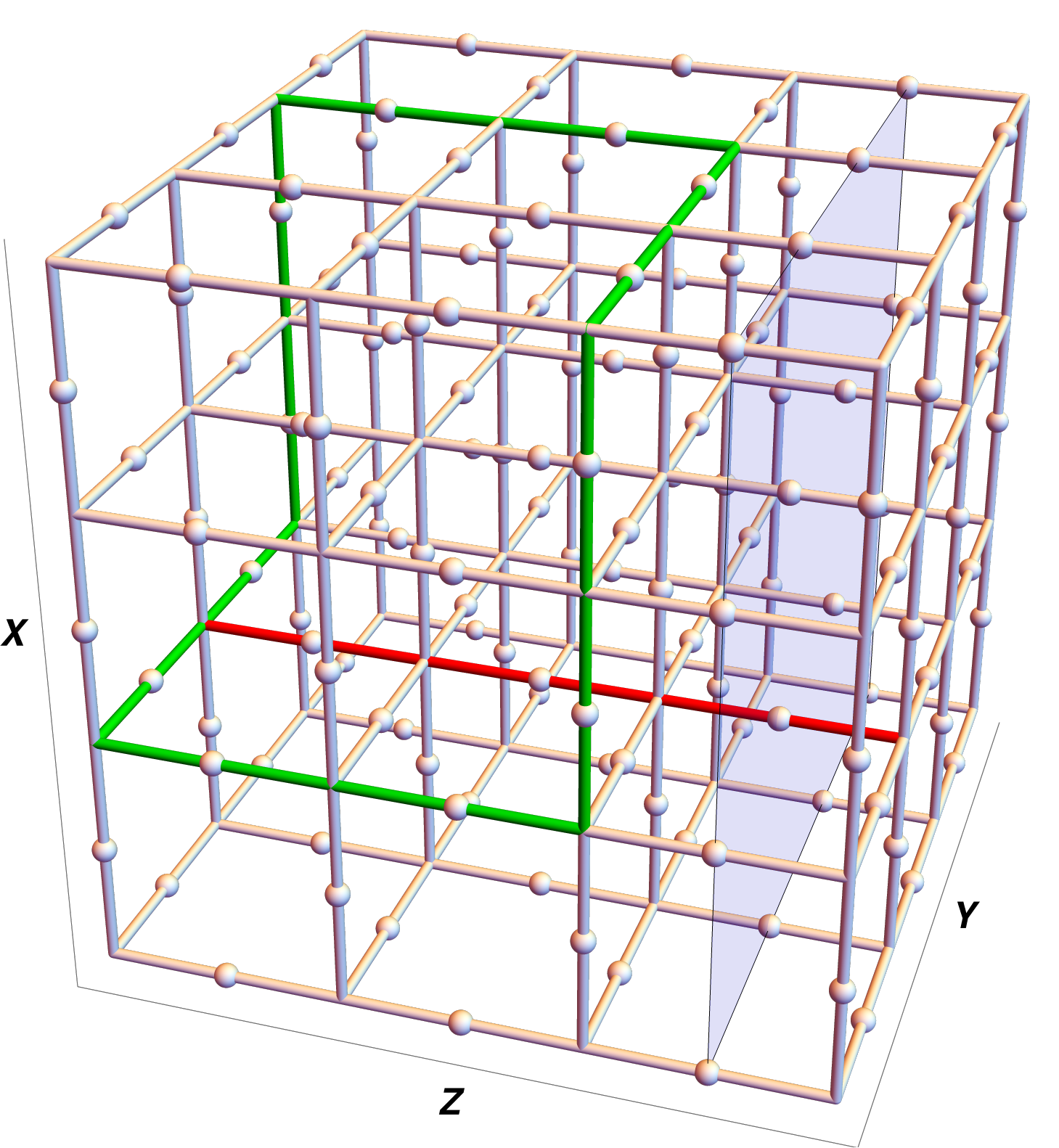}}}
	\caption{{3D pure gauge model}.  A spin $S_\ell$ is placed at each link $\ell$ of the cubic lattice. In the 
		figure it is denoted by a ``sphere". The contour represented in green (light) is a typical Wilson loop. The red (solid) line passing through the links in the periodic $z$-direction represents a Polyakov loop $p$; see details in the text. Flipping all the spins lying on the gray (shaded) plane changes the sign of $p$. Such an operation is referred to as global $\mathbb{Z}_2$ \textit{center} transformation.}
	\label{Sfig}
\end{figure}

In this section, we present the methods used to simulate the model Hamiltonian~(1) of the main text
and some other technical details; see Fig.~\ref{Sfig} for a schematic plot of the system defined on a simple cubic lattice.

We use Monte Carlo methods to generate the equilibrium spin configurations. With them we study the phase transition in the geometrical network formed by the lines piercing the frustrated  plaquettes. 
For this purpose, the initial system configuration at time $t=0$ is prepared in a high temperature phase by assigning 
equally probable random values $(\pm 1)$ to each link spin $S_{\ell}$. Such a configuration belongs to the \textit{confined} phase, where each plaquette term $U_{\rm P}$ of the Hamiltonian~(1) takes a value $+1$ or $-1$ with probability a half. 
The system is then quenched to $39$ different temperature values $T \in [0.8,1.7]$, which lie above, below, and at the critical temperature $T = T_{\rm c}$. We exploit the single spin flip Metropolis rule~\cite{newman1999monte} to evolve the system at the different quench temperatures, with the transition rate
\be
W(S_{\ell} \to -S_{\ell}) = \frac{1}{{\cal N}}\min \left\{1,e^{-\frac{\Delta E}{T}}\right \}
\; ,
\label{metrop}
\ee
where $\Delta E$ is the energy difference generated by the proposed spin flip, the Boltzmann constant $k_{\rm B}$ is set to unity hereafter, and ${\cal N} = 3L^3$ is the number of link spins in the system with $L$ being its linear size. Notice that the flip of a single spin $S_{\ell}$ at link $\ell$ causes the flip of the sign of all four plaquettes sharing the common link $\ell$. This ensures that  gauge invariance is respected during the dynamics. Time is measured in terms of Monte Carlo Steps (MCS), each corresponding to ${\cal N}$ attempted spin flips.

To ensure that the system evolved by the above rule~\eqref{metrop} reaches thermal equilibrium at the quench temperature $T$, we investigate the behavior of average internal energy as a function of time. We let each system of size $L$ thermalize until $t \sim 1\times 10^5$ MCS ($t \sim 5\times 10^5$ MCS at low $T$) after a start from $t=0$. We confirm that the average internal energy saturates to a temperature dependent but time independent value. In order to obtain several equilibrium configurations from a single Monte Carlo run, we further ensure that a configuration measured at time $t$ has become completely decorrelated from a previous equilibrium configuration at time $t_0 < t$. This is done by calculating the spin-spin autocorrelation function $\overline{ S_{\ell}(t_0) S_{\ell}(t) }$, whereby $\overline{ (\ldots) }$ represents a spatial average, i.e., an average over all the $3L^3$ link spins. This quantity decays from unity with increase in time $t$ and finally attains a value around zero when a complete decorrelation from the initial state at $t_0$ has occurred.

Since our primary goal is to investigate the phase transition characteristics of the geometrical networks in an infinitely size system, we undertake a finite size scaling analysis using systems of linear sizes $L = 8, 12, 16, 24, 32, 48, 64$; with each system defined on a simple cubic lattice. The lattice spacing/constant is fixed to unity without the loss of any generality. Additionally, periodic boundary conditions (PBCs) are imposed in all the three spatial directions.

To gain good statistical accuracy, we average the observable quantities of interest (see section ahead) over a large number of equilibrium configurations for a fixed system size ($100 000 - 200 000$ for $L = 8 - 32$, and $50 000$ for $L = 48$ and $L = 64$). This complete study took over a half million of CPU hours.

\section{Thermodynamic phase transition}
\label{ssec2}

This section summarizes the thermodynamic phase transition in the model system~(1) of the main text. 

\begin{figure}[t!]
	\centering
	\rotatebox{0}{\resizebox{.45\textwidth}{!}{\includegraphics{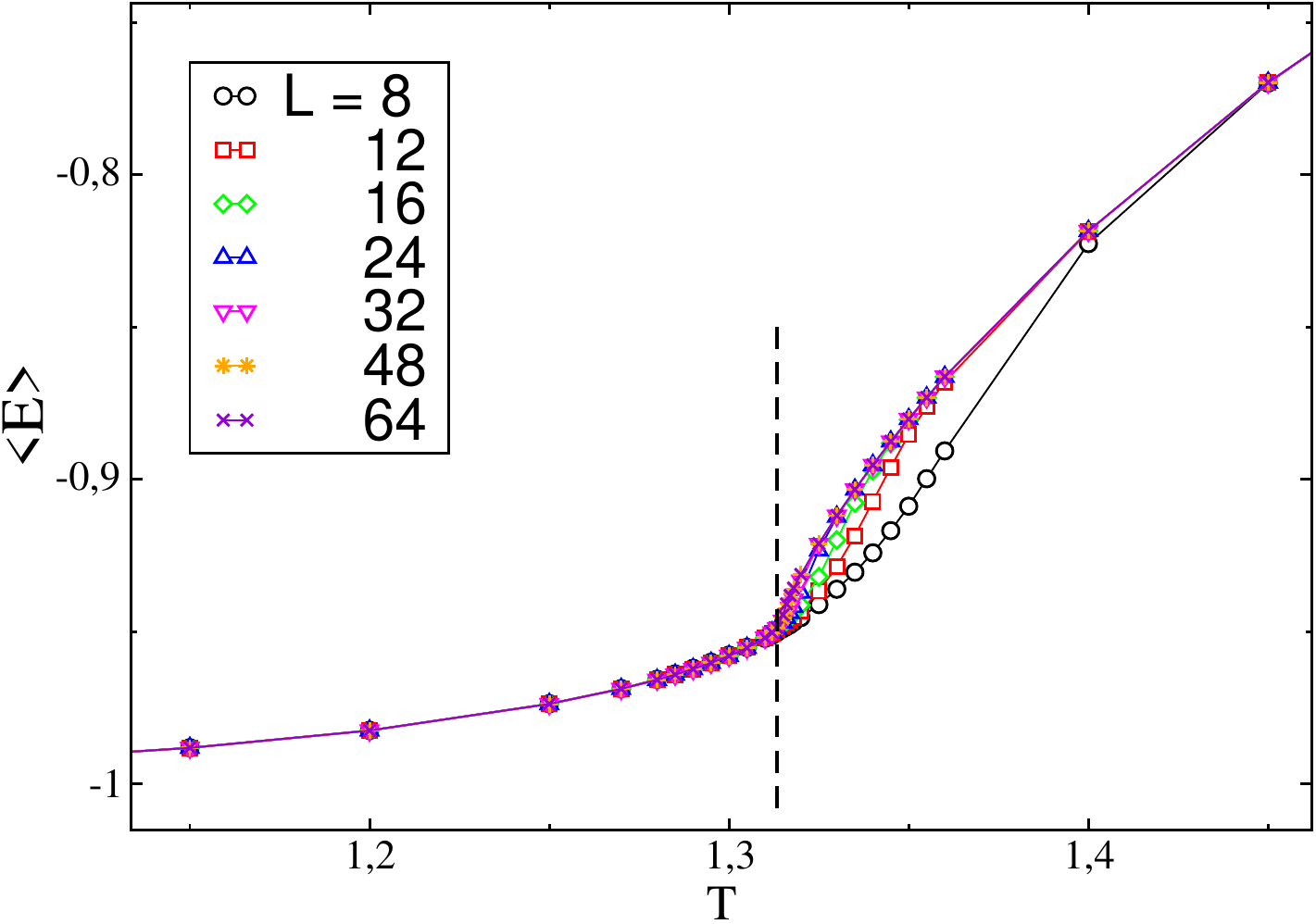}}}
	\caption{Average energy per spin $\langle E \rangle$ vs temperature $T$ for different system sizes (see the key). The dashed vertical line locates the critical temperature $T_{\rm c}$~\cite{10.1063/1.1665530,RevModPhys.51.659}.}
	\label{Sfig1}
\end{figure}

\begin{figure}[t!]
	\centering
	\rotatebox{0}{\resizebox{.45\textwidth}{!}{\includegraphics{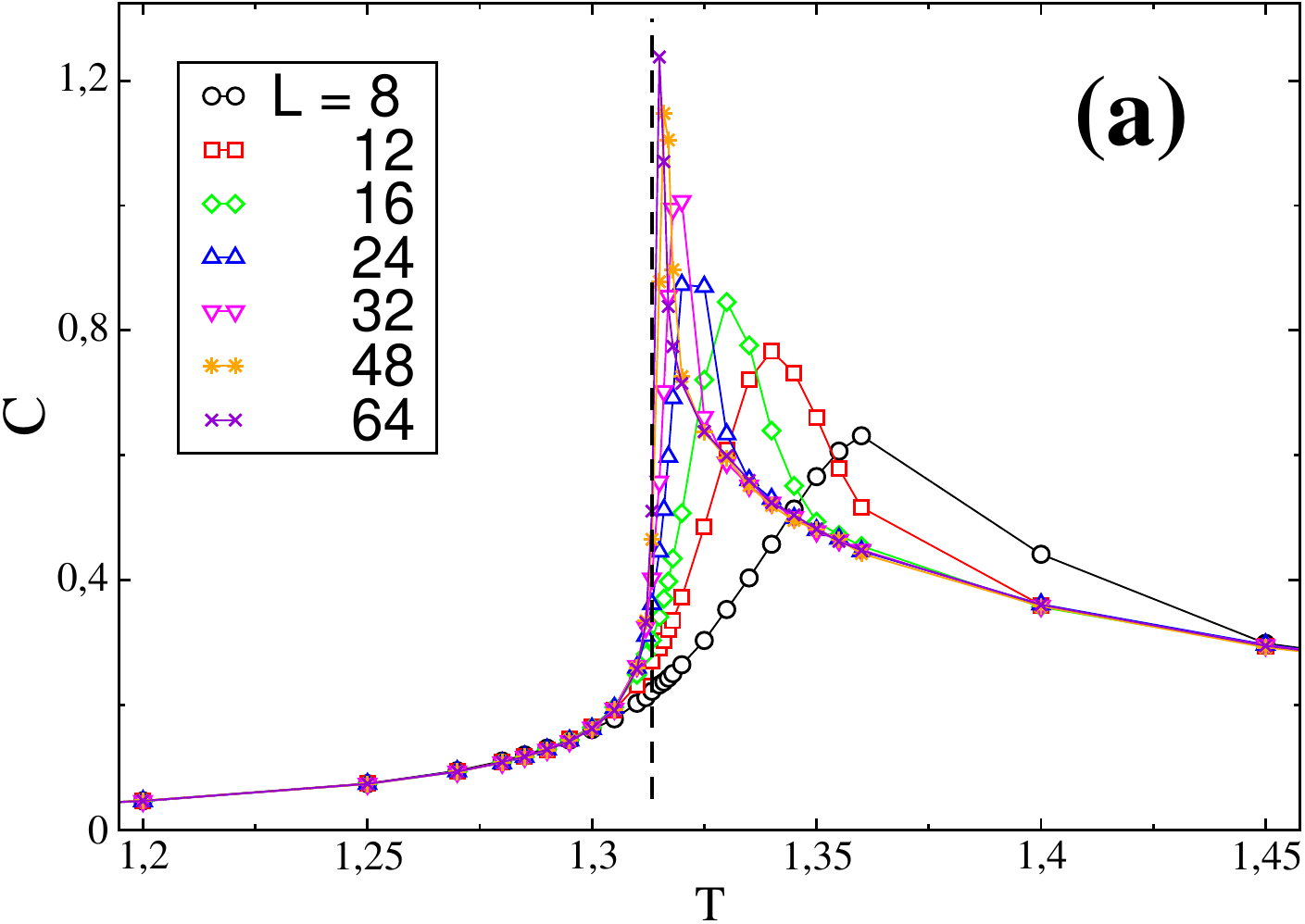}}}
	\rotatebox{0}{\resizebox{.45\textwidth}{!}{\includegraphics{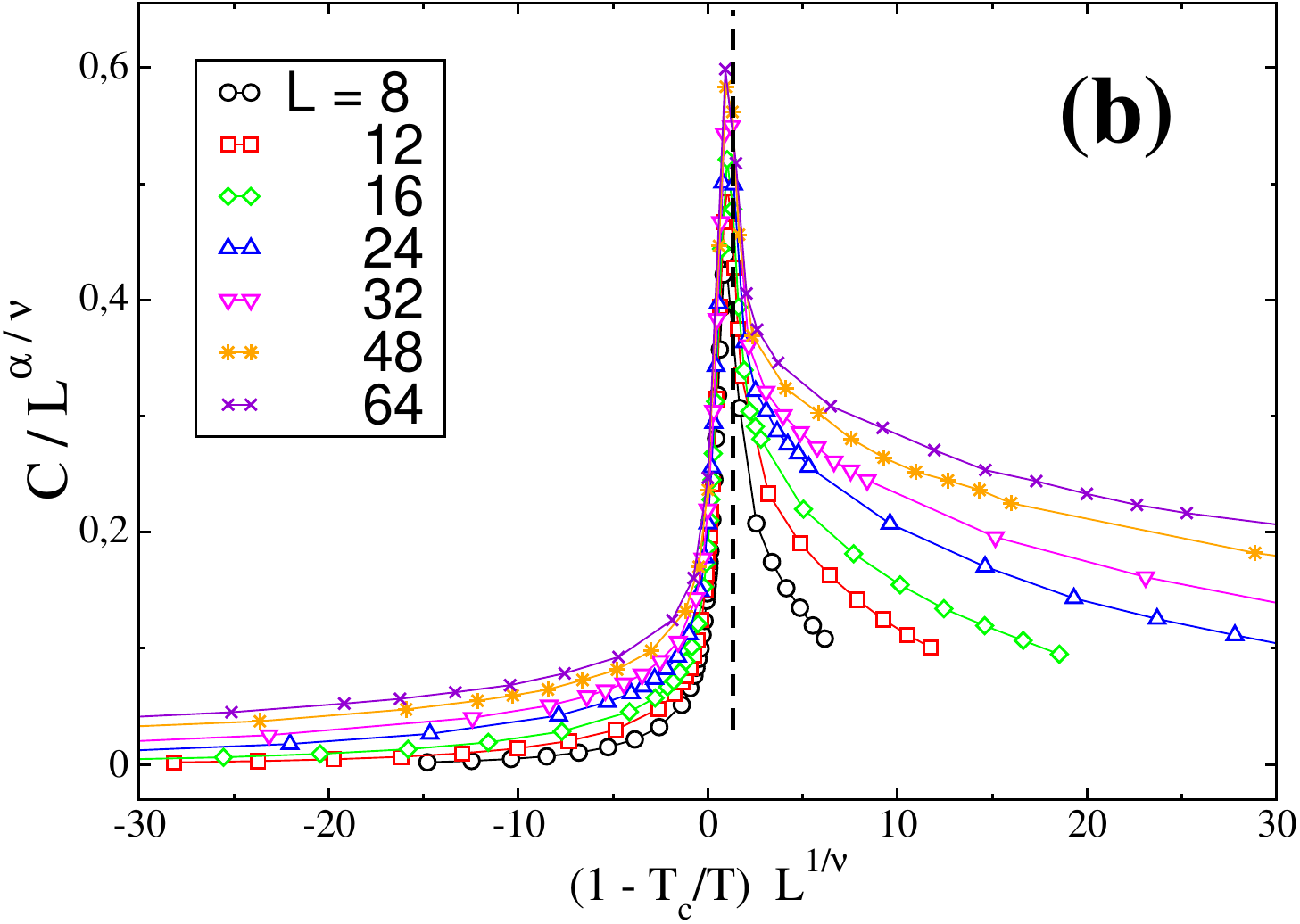}}}
	\caption{(a) Specific heat $C$ vs temperature $T$ for different system sizes (see the key). (b) Scaling plot 
		$C/L^{\alpha/\nu}$ vs $(1-T_{\rm c}/T) L^{1/\nu}$ for the raw data in (a). The dashed vertical line locates  the critical temperature $T_{\rm c}$. The lines joining different symbols are shown as a guide to eye.}
	\label{Sfig2}
\end{figure}

The duality property~\cite{10.1063/1.1665530,RevModPhys.51.659} shows that this model undergoes a phase transition at the critical temperature $T_{\rm c} \simeq 1.313346$ $J/k_{\rm B}$, with the possibility that all thermal critical exponents are identical to those of the dual three-dimensional (3D) Ising model. The magnetization remains zero at all temperatures; however, the average internal energy per spin $\langle E \rangle$ and its fluctuations (i.e., the specific heat $C$) exhibit critical behavior near $T_{\rm c}$. Notice that in the model~(1), $\langle E \rangle$ can be calculated from the average number of plaquettes $\langle {\cal N}_{\rm -} \rangle$ with $-1$ value,
\be
\langle E \rangle = J \left( 2 \frac{\langle {\cal N}_{\rm-} \rangle}{{\cal N}} - 1 \right)
\; ,
\label{Seq00}
\ee
where the symbol $\langle (\ldots) \rangle$ represents the ensemble average, i.e., the average over independent equilibrium configurations. Remember that ${\cal N}$ denotes the number of all plaquettes (both with values $+1$ and $-1$), which is also the same as the number of link spins for our cubic lattice system. In Fig.~\ref{Sfig1}, $\langle E \rangle$ is plotted against $T$ for different system sizes $L$. For increasing values of $L$, the variation of $\langle E \rangle$ with $T$ becomes more and more pronounced near $T_{\rm c}$, as expected. The specific heat can be calculated from the following relation
\be
C = \frac{\langle E^2 \rangle - \langle E \rangle^2}{T^2}
\; , 
\label{Seq01}
\ee
valid in equilibrium. It obeys the finite size scaling form,
\be
C =  L^{\alpha/\nu} F\left[ \left(1-\frac{T_{\rm c}}{T}\right) L^{1/\nu}  \right]
\; ,
\label{Seq02}
\ee
where $\alpha$ and $\nu$ are the standard thermal critical exponents. In Fig.~\ref{Sfig2}(a), $C$ is plotted against temperature $T$. For increasing values of $L$, the peak shifts towards the critical temperature $T_{\rm c}$. The finite size scaling function~\eqref{Seq02} is tested in Fig.~\ref{Sfig2}(b), where we used the known critical exponents of the 3D IM. Clearly, the scaling of datasets is not good at all. It was expected due to strong logarithmic corrections to the power law $C \propto \vert 1 - T_{\rm c}/T \vert ^{-\alpha}$. Notice that the latter holds very close to $T_{\rm c}$ only; otherwise, $C \propto \log \vert 1 - T_{\rm c}/T \vert + \text{const.}$, e.g., see Ref.~\cite{PhysRevE.94.062146}.

In the absence of a local order parameter, gauge invariant correlation functions are used as the global order parameters in model~(1), e.g., Wilson loops~\cite{PhysRevD.10.2445,RevModPhys.51.659},
\be
W(\mathbb{C}) =  \prod_{\ell \in \mathbb{C}} S_{\ell}
\; ,
\label{Seq03}
\ee
where $\mathbb{C}$ is a contour of link spins. The ensemble average of $W(\mathbb{C})$ falls exponentially with the \textit{minimal} area covered by the contour $\mathbb{C}$ in the phase above (confinement) the critical temperature $T_{\rm c}$, while it falls exponentially with the perimeter of $\mathbb{C}$ in the phase below (deconfinement) $T_{\rm c}$, as the loop becomes larger~\cite{RevModPhys.51.659}. It indicates that this quantity is prone to strong finite size effects and not suitable to characterize different phases across $T_{\rm c}$ on a small sized system. A typical construction of a Wilson loop on the cubic lattice is shown in Fig.~\ref{Sfig}.

There is another related quantity, the so-called Polyakov loop $\mathit{p}$~\cite{POLYAKOV197582,SVETITSKY1982423}, which is also interesting. It is simply the product of link spins lying on a line spanning across one fixed periodic boundary direction, say $z$ (see Fig.~\ref{Sfig}). By construction, this quantity remains invariant under the local gauge transformation. However, it is not invariant under a global $\mathbb{Z}_2$ center symmetry~\cite{SVETITSKY1982423}. The spatial average of $\mathit{p}$ over the $xy$-plane, say $\bar{\mathit{p}}$, is zero at high temperatures, and $+1$ or $-1$ in the ground state at $T=0$.

\begin{figure}[t!]
	\centering
	\rotatebox{0}{\resizebox{.45\textwidth}{!}{\includegraphics{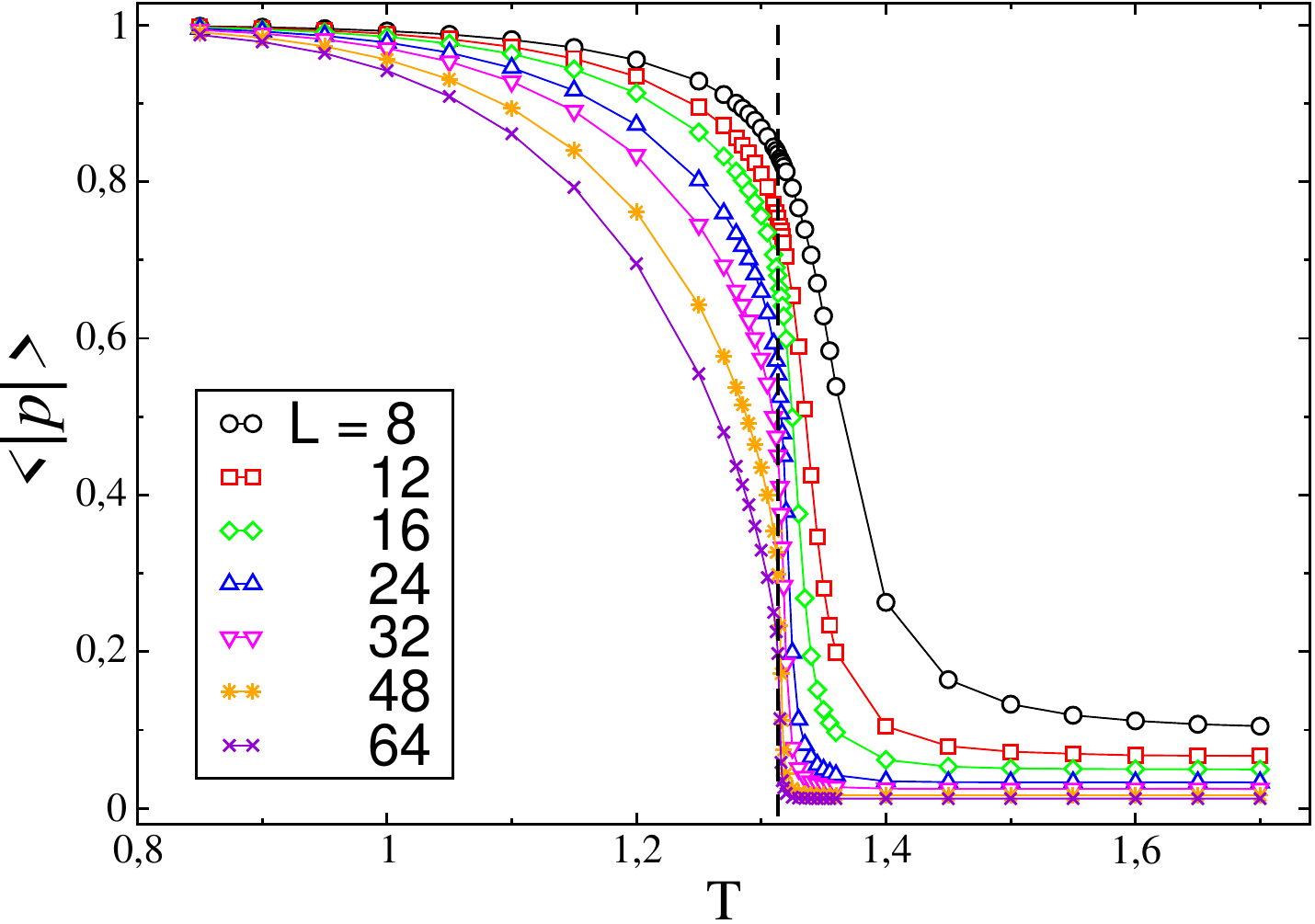}}}
	\caption{Average Polyakov loop $\langle \vert \bar{\mathit{p}} \vert \rangle$ vs temperature $T$ for different system sizes (see the key). The dashed vertical line is placed at the critical temperature $T_{\rm c}$.}
	\label{Sfig3}
\end{figure}

In Fig.~\ref{Sfig3}, the ensemble averaged absolute value of the Polyakov loop $\langle \vert \bar{\mathit{p}} \vert \rangle$ is plotted against temperature $T$ for different system sizes $L$. To our understanding, this quantity is not directly related to any finite size scaling arguments~\cite{KEHL1988324}, and consequently, it cannot be reliably used to extract critical exponents.

\section{Geometric observables}
\label{ssec3}

In this section, we detail different observable quantities used to study the geometrically defined objects that 
we use in the main text. 

Let us first list all the observables before discussing their definitions and scaling properties.
\\
\indent (1) The percolation strength $P$.\\
\indent (2) The susceptibility $\chi$.\\
\indent (3) The Binder cumulant $U_{\rm 4}$.\\
\indent (4) The spanning probability $P_{\rm S}$.\\
\indent (5) The number density $N(s)$ of objects with mass $s$.\\
\indent (6) The average mass $\langle s_{l} \rangle$ of the largest object.\\
\indent (7) The average mass $\langle s \rangle$ of an object with radius of \\
\indent \ \ \ \ \ gyration $R_{\rm g}$.\\
These are standard quantities used in the study of percolation problems~\cite{stauffer2018introduction,Essam_1980,SABERI20151,Strelniker2009}.
We have used the name ``mass'' to indicate the ``size'', that is, the number of elements, in 
an object.

The percolation strength $P$ is the average of the fluctuating mass fraction $m = s_l/(3L^3)$  contributed by the largest geometrical object in a configuration. For an infinite system, this quantity grows as $P \propto (T - T_{\rm p})^{\beta_{\rm p}}$ above the percolation threshold $T_{\rm p}$ and it vanishes otherwise. That is why $P$ plays the role of an order parameter in the percolation transition.

Another closely related quantity which can be fetched from the mass fraction $m$ is the susceptibility $\chi$. It is defined as,
\be
\chi = \frac{L^{D}}{T} \left( \langle m^2 \rangle - \langle m \rangle^2  \right)
\; ,
\label{Seq0}
\ee
where the symbol $\langle (\ldots) \rangle$ denotes the average over independent equilibrium configurations. At $T_{\rm p}$, it diverges algebraically as $\chi \propto \vert T - T_{\rm p} \vert ^{-\gamma_{\rm p}}$.

These quantities obey the following finite size scaling forms,
\bea
P &=& L^{-\beta_{\rm p}/\nu_{\rm p}} f_{P}\left[ \left(1-\frac{T_{\rm p}}{T}\right) L^{1/\nu_{\rm p}}  \right]
\; ,\label{Seqq1}\\
\chi &=& L^{\gamma_{\rm p}/\nu_{\rm p}} f_{\chi}\left[ \left(1-\frac{T_{\rm p}}{T}\right) L^{1/\nu_{\rm p}}  \right]
\; ,
\label{Seq1}
\eea
where $\beta_{\rm p}$, $\gamma_{\rm p}$, and $\nu_{\rm p}$ are the standard critical exponents. The subscript $p$, used here and elsewhere unless explicitly stated otherwise, indicates that the critical exponents are defined from the percolation process. Likewise the thermal phase transitions, one can also calculate the Binder cumulant $U_{\rm 4}$ from the distribution of the mass fraction $m$ of largest object as, $U_{\rm 4} = 1 - \langle m^4 \rangle / 3 \langle m^2 \rangle^2$ (see Sec.~\ref{ssec6} for details).

The spanning probability $P_{\rm S}$ is the fraction of configurations having at least one spanning object (loop/cluster). By spanning object we mean that it crosses/percolates the system in an arbitrary direction. Obviously, in an infinite system, it will be zero below $T_{\rm p}$ and unity above $T_{\rm p}$. Therefore, its scaling form is
\be
P_{\rm S} = f_{S}\left[ \left(1-\frac{T_{\rm p}}{T}\right) L^{1/\nu_{\rm p}}  \right]
\label{Seq2}
\ee
with $\lim_{x\to \infty} f_{S}[x] = 1$ and $\lim_{x\to -\infty} f_{S}[x] = 0$. At $T = T_{\rm p}$, $P_{\rm S}$ attains an $L$-independent \textit{universal} value, $P_{\rm S} = f_{S}[0]$. Due to this characteristic, $P_{\rm S}$ is useful to determine the location of $T_{\rm p}$, whereby the datasets belonging to the different $L$ should intersect each other; see the main text.

The quantity $N(s)$, the number density of loops/clusters with mass $s$ normalized by volume $L^3$, also reveal many interesting properties of thermodynamic as well as geometrical transitions~\cite{stauffer2018introduction,Essam_1980,SABERI20151,Strelniker2009}. In the infinite size limit, below and close to $T_{\rm p}$, $N(s)$ obeys the following form,
\be
N(s)/L^3 \simeq s^{-\tau_{\rm p}} {\rm e}^{-s \epsilon}
\; ,
~~~~
\epsilon \propto ( T_{\rm p} -T ) ^{1/\sigma_{\rm p}}
\; ,
\label{Seq3}
\ee
where $\tau_{\rm p}$ is an exponent related to the interfacial fractal dimension $D_{\rm f}$ of the geometrical objects as, 
\begin{equation}
	\tau_{\rm p} = 1 + D/D_{\rm f}
	\; .
\end{equation}
The  parameter $\epsilon$ is the interfacial tension of the objects such that $\epsilon \to 0$ as $T \to T_{\rm p}$, and the associated exponent is 
\begin{equation}
	\sigma_{\rm p} = 1/(D_{\rm f} \nu_{\rm p})
	\; . 
\end{equation}

Finally, we discuss different methods to calculate the fractal dimension $D_{\rm f}$ of the geometrically defined objects. In an equilibrium configuration at any temperature $T$, the geometric objects are fractal at length scales smaller than the correlation length $\xi$. In particular, at $T_{\rm p}$, $\xi$ is comparable to the system size and, therefore, the average mass $\langle s_{l} \rangle$ of the largest object increases in a power law fashion with the system 
size~\cite{stauffer2018introduction,SABERI20151}
\be
\langle s_{l} \rangle \sim L^{D_{\rm f}}
\; ,
\label{Seq5}
\ee
with $D_{\rm f}$ being the fractal dimension of the object. The notation $\langle (\ldots) \rangle$ was defined above. Taking into account the fact that the average mass satisfies  $\langle s_{l} \rangle \propto L^D P$, then
\begin{equation}
	D_{\rm f} = D - \beta_{\rm p}/\nu_{\rm p}
	\; . 
\end{equation}
Another well-known method to measure a fractal dimension is the box counting approach: plot the average mass $\langle s \rangle$ of an object against its linear extent or radius of gyration $R_{\rm g}$. For fractal objects, 
\be
\langle s \rangle \sim R_{\rm g}^{D_{\rm f}}
\; .
\label{Seq6}
\ee

\section{Low temperature expansion graphs and line loops}
\label{ssec4}

In order to understand the percolation transition of closed line loops formed using the maximal or stochastic connection rules, we look at the low-temperature (LT) expansion of the partition function
\be
\label{Seq7}
{\cal Z} = \sum_{\text{config.}} {\rm e}^{-\beta H} = \sum_{\text{config.}} \prod_{P} {\rm e}^{K U_P}
\; ,
\ee
where $K = \beta J$. Substituting $U_P = 1 - 2 W_P$, we get
\be
\label{Seq8}
{\cal Z} = {\rm e}^{3N K} \sum_{\text{config.}} \prod_{P} {\rm e}^{-2K W_P}
\; .
\ee
Here, $N$ is the number of sites in the system. From the above Eq.~\eqref{Seq8}, one can easily obtain the LT expansion, with ${\rm e}^{-2K}$ being the LT expansion
parameter,
\begin{align}
	\label{Seq9}
	{\cal Z} = {\rm e}^{3N K} 2^{N} \left[\right. 1 &+ 3N \left( {\rm e}^{-2K} \right)^4 \\\nonumber
	&+ 18N \left( {\rm e}^{-2K} \right)^6 + \ldots \left.\right]
	\; .
\end{align}
In the above series, the terms with exponents $4$ and $6$ on the right hand side represent the loop excitations in the ground state, which are formed by the flip of one link-spin and two link-spins belonging to any plaquette, respectively. By flipping more spins in the ground state, higher loop excitations (higher terms of the above series) will be obtained. In this sense, the closed line loops formed by piercing the neighboring frustrated plaquettes (see the main text) are the \textit{LT expansion graphs} of the partition function~\eqref{Seq9} in the loop representation; also see Sec.~\ref{ssec9}.

Now, as we know, the model Hamiltonian~(1) of the main text is (Kramer-Wannier) dual to the 3D pure Ising model. This means that the high-temperature (HT) expansion of the Ising model is analogous to the LT expansion of the gauge model obtained above. The HT expansion of Ising model is as follows:
\begin{align}
	\label{Seq10}
	{\cal Z}^* = \left(\cosh (K^*)\right)^{3N} ~ 2^{N} \left[\right. 1 &+ 3N \left( \tanh (K^*) \right)^4\\\nonumber
	&+ 18N \left( \tanh (K^*) \right)^6 + \ldots \left.\right]
	\; ,
\end{align}
where $K^* = \beta^* J$, and $\tanh (K^*)$ is the HT 
expansion parameter. The series in Eqs.~\eqref{Seq9} and ~\eqref{Seq10} are equal if $\tanh (K^*) = {\rm e}^{-2K}$. The latter happens right at the critical temperature $T = T_c$, i.e., 
\begin{equation}
	\tanh (K^*_c) = {\rm e}^{-2K_c}
	\; .
\end{equation} 
This equation yields the critical temperature of the plaquette model, $T_c$, in terms of the 
critical temperature of the Ising model, $T_{\rm c}^*$, and {\it vice versa}.

The HT graph or loop representation of the Ising model was earlier studied by Winter and co-workers~\cite{PhysRevE.77.061108} using Monte Carlo methods. The idea was to integrate out the spin variables and express the partition function in terms of closed geometrical loops. This is similar to the random cluster representation, where spin variables are integrated out and the partition function is written in terms of occupied/empty bonds; see Sec.~\ref{ssec9}. The closed loops are the HT graphs of different degrees (number of occupied bonds). Shimada et. al~\cite{shimada2016fractal} and Kompaniets et. al~\cite{PhysRevE.101.012104} have also studied the fractal dimensions of these objects using conformal bootstrap and the $D=4-\epsilon$ expansion technique, respectively. It was found~\cite{PhysRevE.77.061108} that in the loop representation of the ${\cal O}(N =1)$ model, the HT graphs, which are also closed loops, show a percolation transition at the Ising critical temperature $T_{\rm c}^*$. At high $T$, there are only a few such small loops on the lattice, while as $T$ decreases the loops become larger both in size and number. At $T < T_{\rm c}^*$, such loops engulf the lattice, with $T_{\rm c}^*$ as their percolation temperature. We also add that in this loop representation the correlation length exponent $1/\nu_{\rm p}$ is related to the scaling dimension $\Delta_\epsilon$ of the energy operator $\epsilon$ in the 3D ${\cal O}(N)$ singlet sector (S). However, the exponent $\beta_{\rm p}/\nu_{\rm p}$ or the fractal dimension $D_{\rm f}$ rather pertain to the scaling dimension $\Delta_T$ of a \textit{nontrivial} symmetric tensor operator $\varphi_{ab}$ in the 3D ${\cal O}(N)$ tensor sector (T). See Ref.~\cite{shimada2016fractal} for details. The latter operator is missing in the Ising Wilson CFT which is a single component theory and ideal for spin representation only. For the loop representation an appropriate theory is obtained by analytic continuation of the order parameter components $N$ to the upper neighborhood of $N=1$, i.e., $N\to 1^+$.

Due to the equivalence of the series~\eqref{Seq9} and \eqref{Seq10}, one expects that the properties of the LT expansion graphs of the gauge model, i.e., closed line loops, should be those of the Ising HT graphs.

\section{Construction of Fortuin-Kasteleyn clusters}
\label{ssec9}

The random cluster representation~\cite{FORTUIN1972536,Coniglio_1980} was introduced initially to understand the phase transition in the Potts model, where the spin variables are integrated out and the bond variables (occupied/empty) are introduced in the partition function. The distinct sets of connected occupied bonds form clusters, known as Fortuin-Kasteleyn (FK) clusters. In the Ising model, which is a special case of Potts model, the partition function is written as follows,
\bea
\label{ssec91}
{\cal Z}^* 
&=& \sum_{\text{config.}} \prod_{<i,j>} {\rm e}^{K^*} \left[ (1-p) + p \delta_{s_is_j,1} \right]
\; ,\\ \nonumber
&=& \!\! \sum_{\text{config.}} \prod_{<i,j>} \sum_{n_{\rm ij} = 0,1} \! {\rm e}^{K^*} \left[ (1-p) \delta_{n_{\rm ij},0} + p \delta_{s_is_j,1} \delta_{n_{\rm ij},1} \right]
\, ,
\;\;\;\;
\eea
where $K^* = \beta^*J$, $p = 1 - {\rm e}^{-2K^*}$, and $s_i$ is an Ising spin sitting at each site of the cubic lattice. $n_{\rm ij}=0,1$ is a bimodal random variable assigned to each bond of the configuration. The bond with $n_{\rm ij} = 1$ is said to be occupied and it is empty if $n_{\rm ij} = 0$. By integrating out the spin variables the partition function can be further expressed as,
\be
\label{ssec92}
{\cal Z}^*_{\rm FK} = \! {\rm e}^{{\cal N} K^*} \! \! \sum_{\rm \{{\cal G}\}:{\cal G}\subseteq {\cal L}^*} \! p^{{\cal N}_{\rm b}^*} (1-p)^{{\cal N}-{\cal N}_{\rm b}^*} ~2^{{\cal N}_{\cal G}^*}
\, ,
\;\;\;\;
\ee
where ${\cal N} (=3N)$ is the total number of bonds in the lattice ${\cal L}^*$, ${\cal G}$ stands for the set of occupied bonds which contains ${\cal N}_{\cal G}^*$ number of connected components, i.e, FK clusters. ${\cal N}_{\rm b}^*$ and ${\cal N}-{\cal N}_{\rm b}^*$ denote the number of occupied bonds and empty bonds in such a graph, respectively. At the (inverse) temperature $\beta^* \le \beta^*_{\rm c}$ (critical one), the FK clusters percolate and their critical exponents are the same as the thermal critical exponents~\cite{Coniglio_1980,DOTSENKO1995577}.

We have seen in Sec.~\ref{ssec4} that in the loop representation the Ising partition function ${\cal Z}^*$ takes the form,
\be
\label{ssec93}
{\cal Z}^*_{\rm loop} =  (\cosh K^*)^{\cal N} ~2^{N} \!\!\!\! \sum_{\text{closed loops}} \!\! g_{\rm \ell}^* ~( \tanh K^* )^{\ell}
\, ,
\;\;\;\;
\ee
obtained via the high temperature expansion of ${\cal Z}^*$. Here, the summation is over all the closed loops with even number $\ell$ of links ($\ell = 0, 2, 4, \ldots$) on ${\cal L}$, and $g_{\rm \ell}^*$ is the degeneracy of loops having $\ell$ number of links. One can also obtain the loop configuration~\eqref{ssec93} directly from an FK one~\eqref{ssec92}, using the Kirchhoff's conservation criterion.

Similarly, we define the random cluster representation or FK clusters for the present gauge model. Such an approach was earlier~\cite{ben1990critical} exploited to obtain a cluster Monte Carlo algorithm. Likewise the Ising model, we assign a bimodal random variable $n_{\rm P}=0,1$ to each plaquette of the lattice. Then, the partition function ${\cal Z} = \sum_{\text{config.}} \prod_{\rm P} {\rm e}^{K U_{\rm P}}$ can be written as,
\bea
\label{ssec94}
{\cal Z}
&=& \sum_{\text{config.}} \prod_{P} {\rm e}^{K} \left[ (1-q) + q \delta_{U_{\rm P},1} \right]
\; ,\\ \nonumber
&=& \!\! \sum_{\text{config.}} \prod_{\rm P} \sum_{n_{\rm P} = 0,1} \! {\rm e}^{K} \left[ (1-q) \delta_{n_{\rm P},0} + q \delta_{U_{\rm P},1} \delta_{n_{\rm P},1} \right]
\, ,
\;\;\;\;
\eea
with $K = \beta J$ and $q \equiv 1 - {\rm e}^{-2K}$. A plaquette is said to be active if $n_{\rm P} = 1$ and it is empty if $n_{\rm P} = 0$. Notice that in the above representation a frustrated plaquette $(U_{\rm P} = -1)$ is always empty, while an unfrustrated 
plaquette $(U_{\rm P} = 1)$ is empty with a probability $1-q$. 
An FK cluster is formed by the neighboring ``empty" plaquettes. The partition function is further written as,
\be
\label{ssec95}
{\cal Z}_{\rm FK} =  
{\rm e}^{{\cal N} K}  \sum_{{\rm G}\subseteq {\cal L}}  q^{{\cal N}_{\rm P}} (1-q)^{{\cal N}-{\cal N}_{\rm P}} ~2^{{\cal N}_{\rm G}}
\, ,
\;\;\;\;
\ee
where ${\cal N} (=3N)$ is the total number of plaquettes in the lattice ${\cal L}$, $G$ is the set of occupied plaquettes, and ${\cal N}_{\rm G}$ denotes the number of connected components which are FK clusters. ${\cal N}_{\rm P}$ and ${\cal N}-{\cal N}_{\rm P}$ represent the number of occupied and empty plaquettes in such a graph, respectively. The loop representation~\eqref{Seq9} of ${\cal Z}$ is
\be
\label{ssec96}
{\cal Z}_{\text{loop}} = {\rm e}^{{\cal N} K} 2^{N} \sum_{\text{closed loops}} g_{\rm \ell} ~ \left( {\rm e}^{-2K} \right)^{\ell}
\; ,
\ee
which can also be obtained from the FK one~\eqref{ssec95} by enforcing the gauge invariance (even number of plaquettes to be empty/occupied in a cube) onto FK graphs.

\section{Geometric loops with maximal rule and FK clusters}
\label{ssec11}

\begin{figure}[h!]
	\vspace{0.5cm}
	\centering
	\rotatebox{0}{\resizebox{.48\textwidth}{!}{\includegraphics{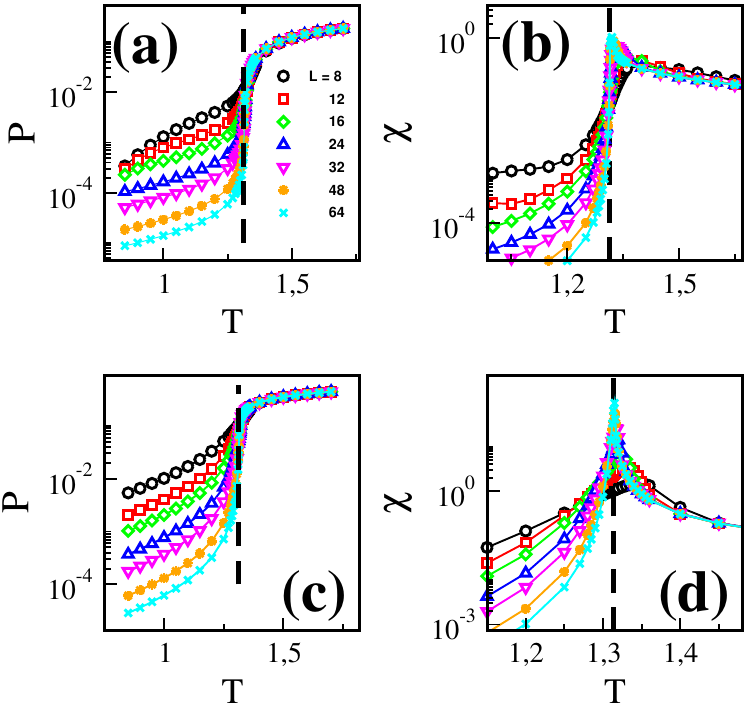}}}
	\caption{(a) and (c) Percolation strength $P$ vs $T$. (b) and (d) Susceptibility $\chi$ vs $T$. Panels (a)-(b) present data
		for geometric loops constructed with the maximal connection rule, while (c)-(d) do for FK clusters. Different datasets represent systems of different linear size $L$, see the key in (a). The vertical dotted lines is located at $T_{\rm c} = 1.3133$.}
	\label{Sfig13}
\end{figure}

Figure~\ref{Sfig13} show some plots of the percolation strength $P$ and the susceptibility $\chi$ for the geometric loops built with the maximal connection rule and the FK clusters. These data complement  the scaling plots of the same quantities in the main text. 
See the caption for details.

\section{Geometric line loops with the stochastic connection rule}
\label{ssec5}

This section discusses some results for the geometric line loops constructed with the stochastic connection rule, which are complementary to those presented in the main text for the geometric line loops built with the maximal connection rule.

\begin{figure}[t!]
	\centering
	\rotatebox{0}{\resizebox{.48\textwidth}{!}{\includegraphics{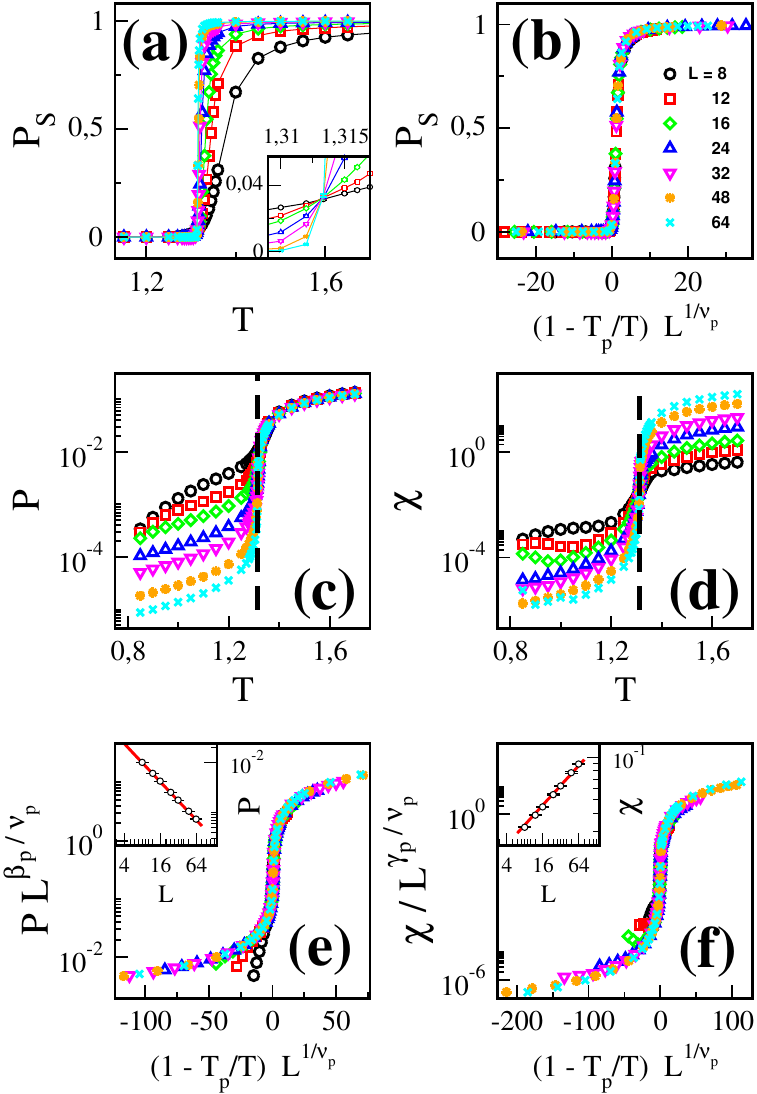}}}
	\caption{Geometric line loops constructed with the stochastic connection rule. (a) Spanning probability $P_{\rm S}$ vs temperature $T$. (b) $P_{\rm S}$ against the scaling variable $(1-T_{\rm p}/T) L^{1/\nu_{\rm p}}$. (c) Percolation strength $P$ vs $T$. (d) Susceptibility $\chi$ vs $T$. (e) Rescaled plot of percolation strength, $P L^{\beta_{\rm p}/\nu_{\rm p}}$ vs $(1-T_{\rm p}/T) L^{1/\nu_{\rm p}}$. (f) Rescaled plot of susceptibility, $\chi / L^{\gamma_{\rm p}/\nu_{\rm p}}$ vs $(1-T_{\rm p}/T) L^{1/\nu_{\rm p}}$. Different datasets represent systems of different linear size $L$, see the key in (b). The inset in (a) magnifies the intersection region in the main frame. The insets in (e) and (f) display the values of $P$ and $\chi$ at the critical temperature $T_{\rm c}$ against the system size $L$, with solid lines representing the laws $P \sim L^{-\beta_{\rm p}/\nu_{\rm p}}$ and $\chi \sim L^{\gamma_{\rm p}/\nu_{\rm p}}$, respectively. The vertical dashed lines in (c)-(d) are located at $T_{\rm c}$. Different parameters are fixed to $T_{\rm p} = 1.3133$, $1/\nu_{\rm p} = 1.58$, $\beta_{\rm p}/\nu_{\rm p} = 1.26$, and $\gamma_{\rm p}/\nu_{\rm p} = 0.47$. The collapses in (b), (e), and (f) remain valid for the numerical estimates of parameters within error bars; see the text.}
	\label{Sfig12}
\end{figure}

In Fig.~\ref{Sfig12}, we plot several quantities that characterize the percolation transition. In (a) we plot the spanning probability $P_{\rm S}$ against temperature for different systems of linear sizes $L$. Similar to what we found with the maximal rule, $P_{\rm S}$ approaches zero for low $T$, while it tends to unity for high $T$. The value of the percolation temperature $T_{\rm p}$ obtained from the intersection point of the datasets for different $L$ (see inset) is $T_{\rm p} \simeq 1.3132(2)$, which is in excellent agreement with the critical temperature $T_{\rm c}$ (obtained from the duality relation) as well as with the one obtained from the maximal connection rule. In (b) $P_{\rm S}$ is plotted against the scaling variable $(1-T_{\rm p}/T) L^{1/\nu_{\rm p}}$ to verify the scaling relation~\eqref{Seq2}. Here, we fix $T_{\rm p} = T_{\rm c}$ as obtained in panel (a) and we treat the exponent $\nu_{\rm p}$ as a free parameter. An excellent scaling collapse is observed for $1/\nu_{\rm p} \simeq 1.60(3)$, which also matches quite well with $1/\nu_{\rm p} \sim 1.59$ for the 3DIM within numerical precision.

\begin{figure}[t!]
	\centering
	\rotatebox{0}{\resizebox{.48\textwidth}{!}{\includegraphics{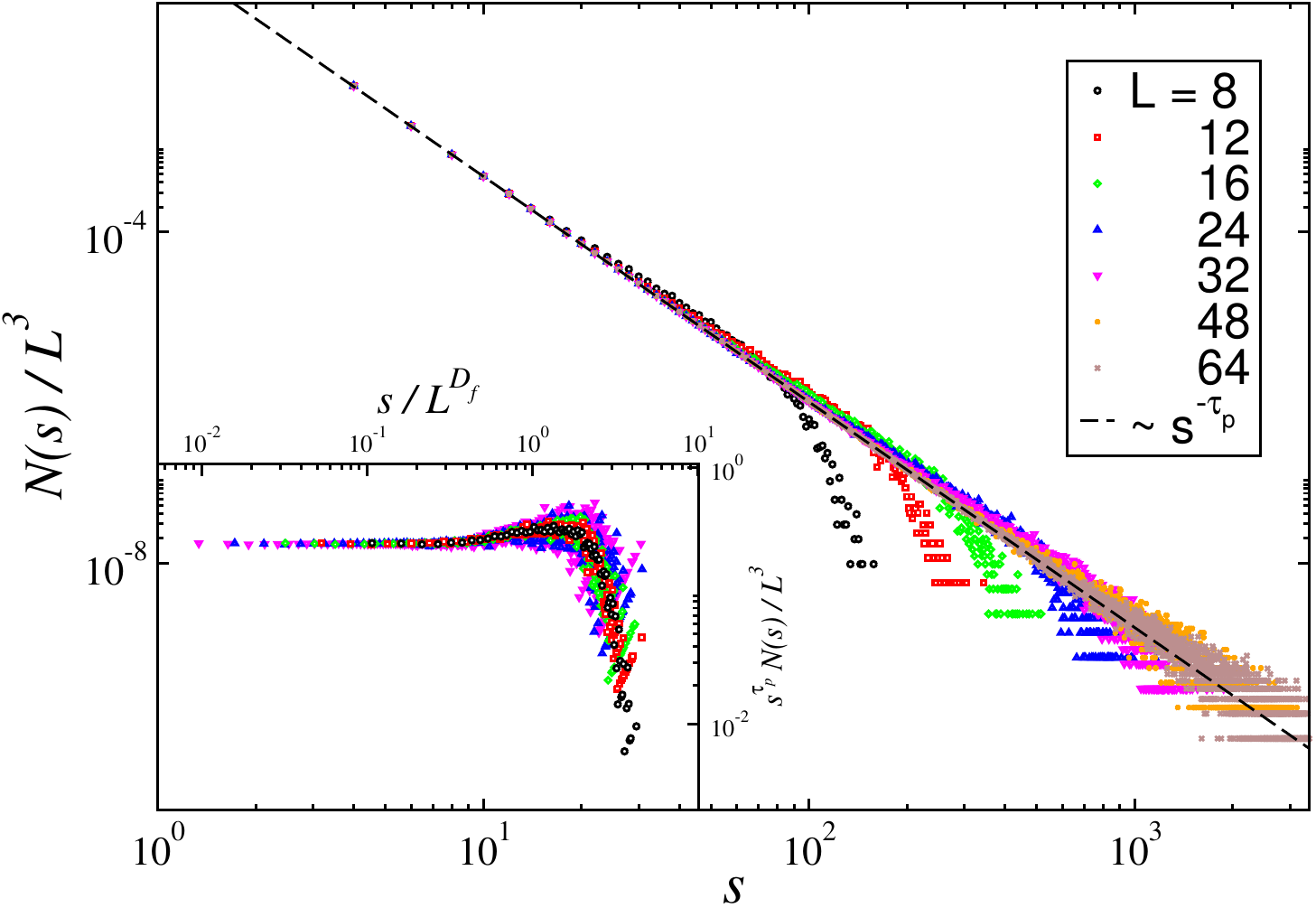}}}
	\caption{Geometric line loops with the stochastic connection rule. Plots of $N(s)/L^3$ vs mass $s$ for different system sizes $L$ (see the key) at the critical temperature $T_{\rm c}$. The dashed line represents the law $N(s)/L^3 \propto s^{-\tau_{\rm p}}$, with $\tau_{\rm p} = 2.71$. The inset plots the scaling variable $s^{\tau_{\rm p}} N(s)/L^3$ against $s/L^{D_{\rm f}}$ for data in the main frame. Herein, $\tau_{\rm p}$ and $D_{\rm f}$ are fixed to $\tau_{\rm p} = 2.71$ and $D_{\rm f} = 1.74$.}
	\label{Sfig5}
\end{figure}

In panels (c) and (d) of Fig.~\ref{Sfig12}, the percolation strength $P$ and susceptibility $\chi$ are plotted against $T$, respectively. While the behavior of $P$ is akin to the one of the loops built with the maximal connection rule, $\chi$ shows a strange behavior for $T \ge T_{\rm c}$; see Sec.~\ref{ssec11} for comparison with the results for the maximal connection rule. Generally, a peak in $\chi$ is observed around $T_{\rm c}(L)$, which approaches the theoretical value of $T_{\rm c}$ with $L$ (i.e., in the thermodynamic limit). Beyond $T > T_{\rm c}(L)$ the $\chi$ continuously decreases with $T$. However, in the present panel (d) $\chi$ seems to increase. The reason for this unexpected behavior is the stochastic connection method of geometric loops~\cite{PhysRevE.94.062146,Kobayashi_2016}. In this method, whenever a branch point in the line loops is encountered, a dice (random variable between $0$ and $1$) is thrown to resolve it such that the loops do not cross each other. Therefore, the length of the resulting loops is shorter than the length of those obtained from the maximal connection methods. Especially, above $T_{\rm c}$ as the number of frustrated plaquettes increases with $T$, the number of loops obtained from the stochastic rule also increases. Therefore, the susceptibility $\chi$ (variance of mass fraction $m$ of the largest loop in the system, see Eq.~\eqref{Seq0}), also increases with $T$. The latter does not happen for the maximal connection rule as only a few loops evade the whole system and only very small other loops complete the picture.

In panels (e) and (f) of the same figure, the scaling functions [Eqs.~\eqref{Seqq1}-\eqref{Seq1}] of P and $\chi$ are checked, and the power-law behavior of these quantities at $T_{\rm c}$ is verified in the respective insets. Both main panels show excellent data collapse when using $T_{\rm p} = T_{\rm c}$ and the value of $1/\nu_{\rm p}$ as obtained above, and treating the exponents $\beta_{\rm p}/\nu_{\rm p}$ and $\gamma_{\rm p}/\nu_{\rm p}$ as free parameters. The values of these exponents so obtained agree with the power law behavior of the quantities $P \sim L^{-\beta_{\rm p}/\nu_{\rm p}}$ and $\chi \sim L^{\gamma_{\rm p}/\nu_{\rm p}}$ in the insets, respectively [$\beta_{\rm p}/\nu_{\rm p} \simeq 1.269(7)$ and $\gamma_{\rm p}/\nu_{\rm p} \simeq 0.478(16)$]. Interestingly, the scaling collapse of $\chi$ remains good in the full range of $T$, despite the fact that the behavior of $\chi$ is unusual for $T>T_{\rm c}$.

Finally, akin to the maximal rule, for the stochastic rule the values of exponents $1/\nu_{\rm p}$, $\beta_{\rm p}/\nu_{\rm p}$ and $\gamma_{\rm p}/\nu_{\rm p}$ are the same as those of HT graphs in 3DIM within numerical precision.

In Fig.~\ref{Sfig5}, the number density of loops, built with the stochastic rule, with mass $s$ at $T=T_{\rm c}$ is plotted. The power law behavior~\eqref{Seq3} is clearly obeyed in the plot, with the exponent $\tau_{\rm p} \sim 2.72(1)$ which is in good agreement with the one found with the maximal rule; see main text. In the inset the scaling relation~\eqref{Seq11} is also verified.

\section{Binder cumulant of the percolation order parameter}
\label{ssec6}

\begin{figure}[t!]
	\centering
	\rotatebox{0}{\resizebox{.4\textwidth}{!}{\includegraphics{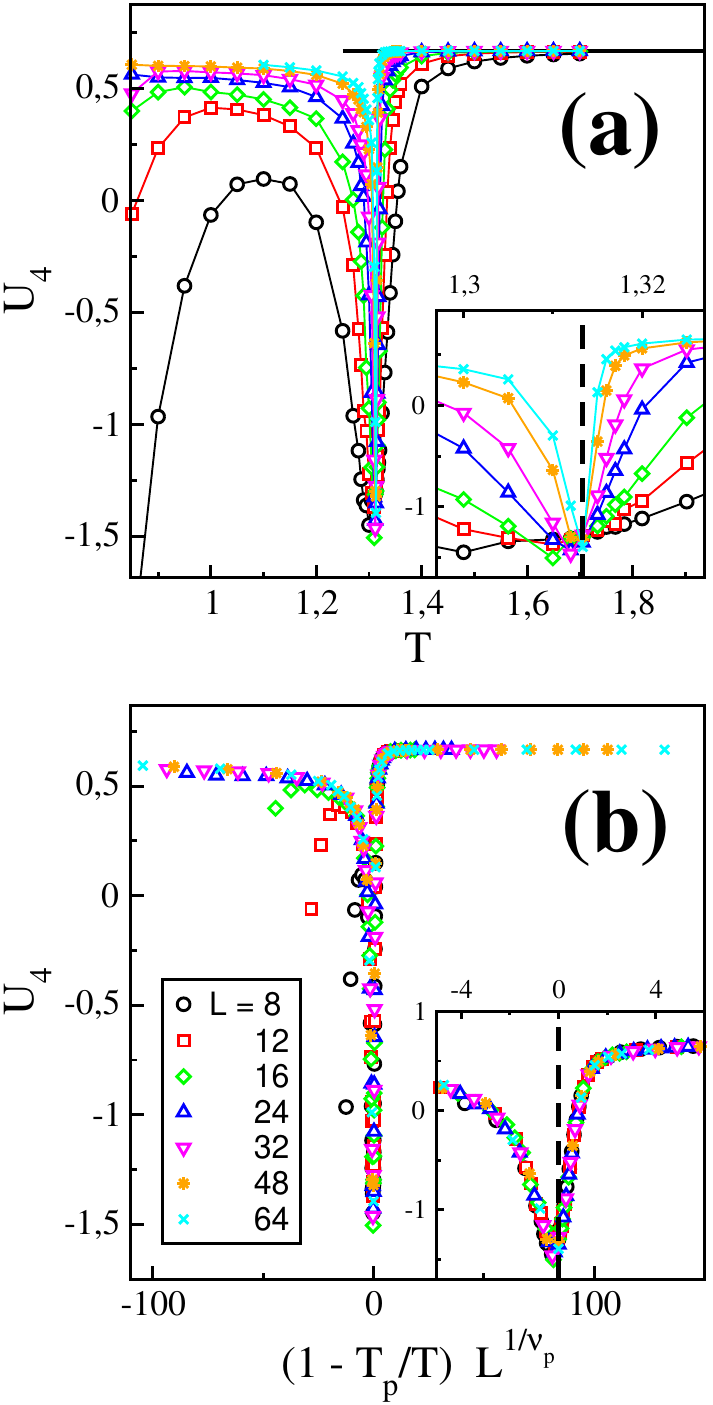}}}
	\caption{(a) Binder cumulant $U_{\rm 4}$ vs temperature $T$ for line loops formed with the maximal connection rule. The horizontal line represents the large-$T$ behavior, i.e, $U_{\rm 4} = 2/3$. The inset in (a) magnifies the dip and intersection region, with the vertical dashed line locating the critical temperature $T_{\rm c}$. (b) Scaling plot of $U_{\rm 4}$ for the data in (a), having fixed $T_{\rm p} = 1.3133$ and $1/\nu_{\rm p} = 1.58$. The inset in (b) magnifies the data collapse of the main frame around the dip. Different datasets represent the systems of different linear size $L$; see the key in (b).}
	\label{Sfig5a}
\end{figure}

The Binder cumulant $U_{\rm 4}$ is an important quantity to characterize the nature of geometric as well as thermal phase transitions~\cite{PhysRevLett.47.693,PhysRevB.30.1477}. In the present system, it can be calculated from the distribution $P_{\text{dist}}(m)$ of the mass fraction $m$ of largest object as,
\be
\label{Seq22}
U_{\rm 4} = 1 - \frac{\langle m^4 \rangle}{3 \langle m^2 \rangle^2}
\; ,
\ee
where the symbol $\langle (\ldots) \rangle$ represents the average over independent equilibrium configurations. The ratio $\langle m^4 \rangle / \langle m^2 \rangle^2$ represents the kurtosis of the distribution $P_{\text{dist}}(m)$. In the percolation phase $T > T_{\rm p}$, both $\langle m^4 \rangle$ and $\langle m^2 \rangle^2$ are equal, i.e, $U_{\rm 4} = 2/3$ for $T \gg T_{\rm p}$. However, at low $T$, the geometrical objects are scarce as very less number of plaquettes are frustrated (no frustrated plaquette at $T=0$). As a result, the distribution of $m$ becomes skinny around the mean with long tails.

In a continuous phase transition, $U_{\rm 4}$ has a scaling form analogous to the one of the spanning probability~\eqref{Seq2}. It reads
\be
\label{Seq23}
U_{\rm 4} = f_{U}\left[ \left(1-\frac{T_{\rm p}}{T}\right) L^{1/\nu_{\rm p}}  \right]
\; .
\ee
At $T = T_{\rm p}$, $U_{\rm 4}$ takes a universal value independent of system size $L$, $U_{\rm 4} = f_{U}[0]$, i.e., the datasets belonging to the different system size should intersect each other.

\begin{figure}[h!]
	\centering
	\rotatebox{0}{\resizebox{.4\textwidth}{!}{\includegraphics{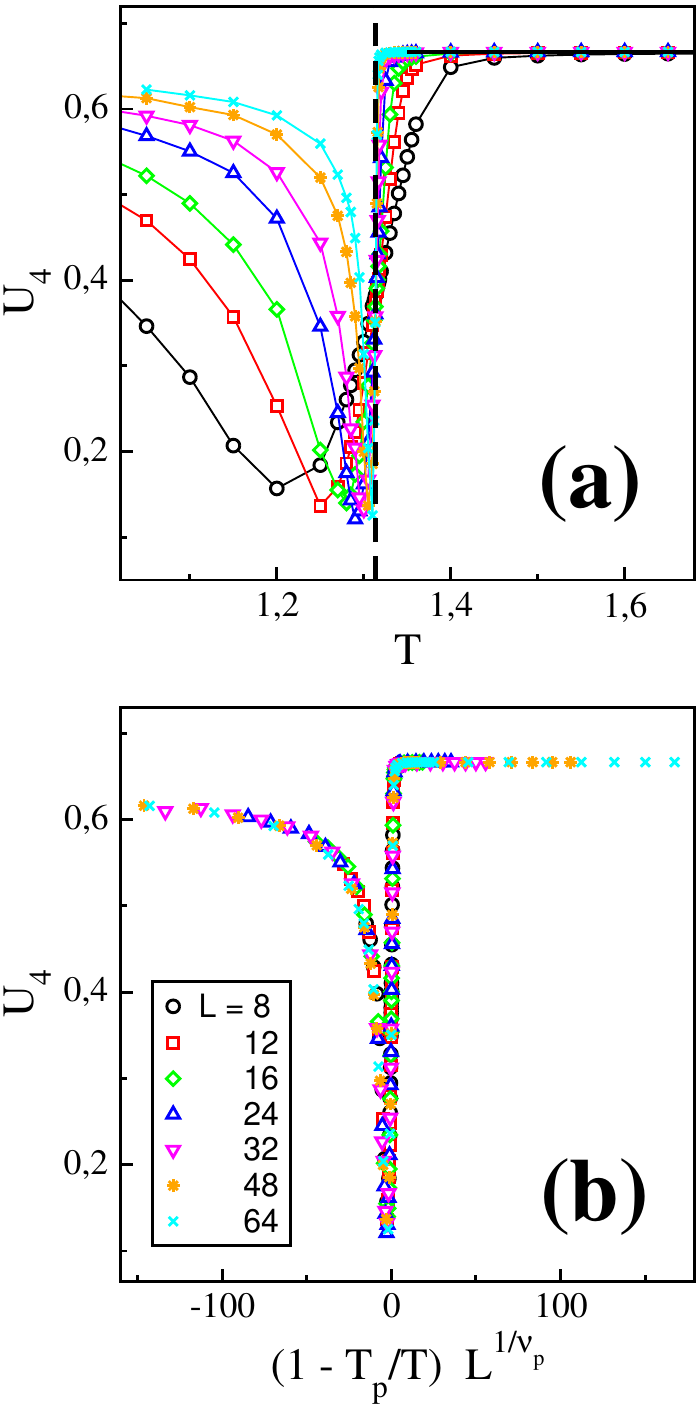}}}
	\caption{Plots similar to Fig.~\ref{Sfig5a}, but for FK graphs. In (b), the parameters are fixed to $T_{\rm p} = 1.3133$ and $1/\nu_{\rm p} = 1.58$.}
	\label{Sfig5b}
\end{figure}

\begin{figure}[t!]
	\centering
	\rotatebox{0}{\resizebox{.48\textwidth}{!}{\includegraphics{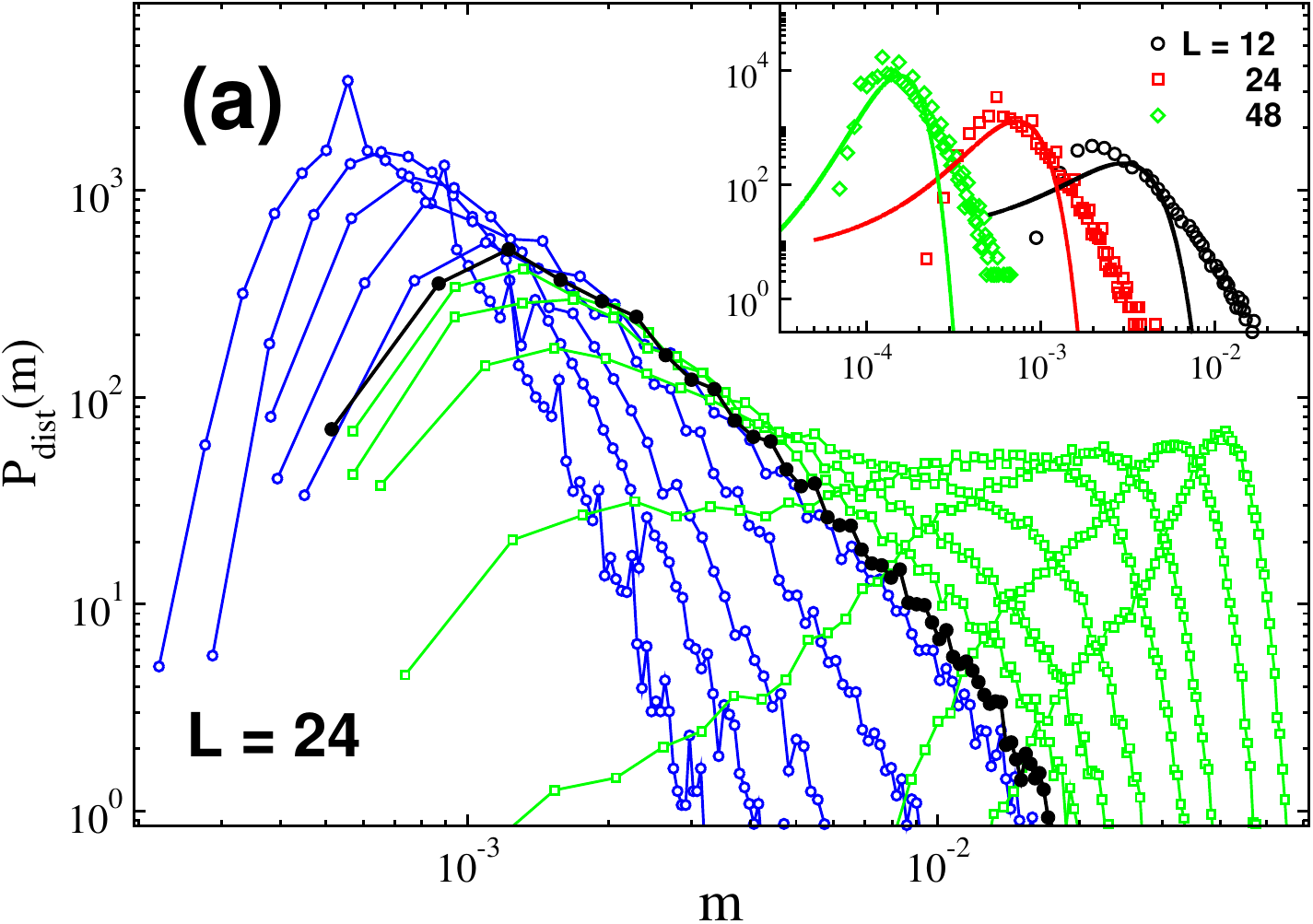}}}
	\rotatebox{0}{\resizebox{.48\textwidth}{!}{\includegraphics{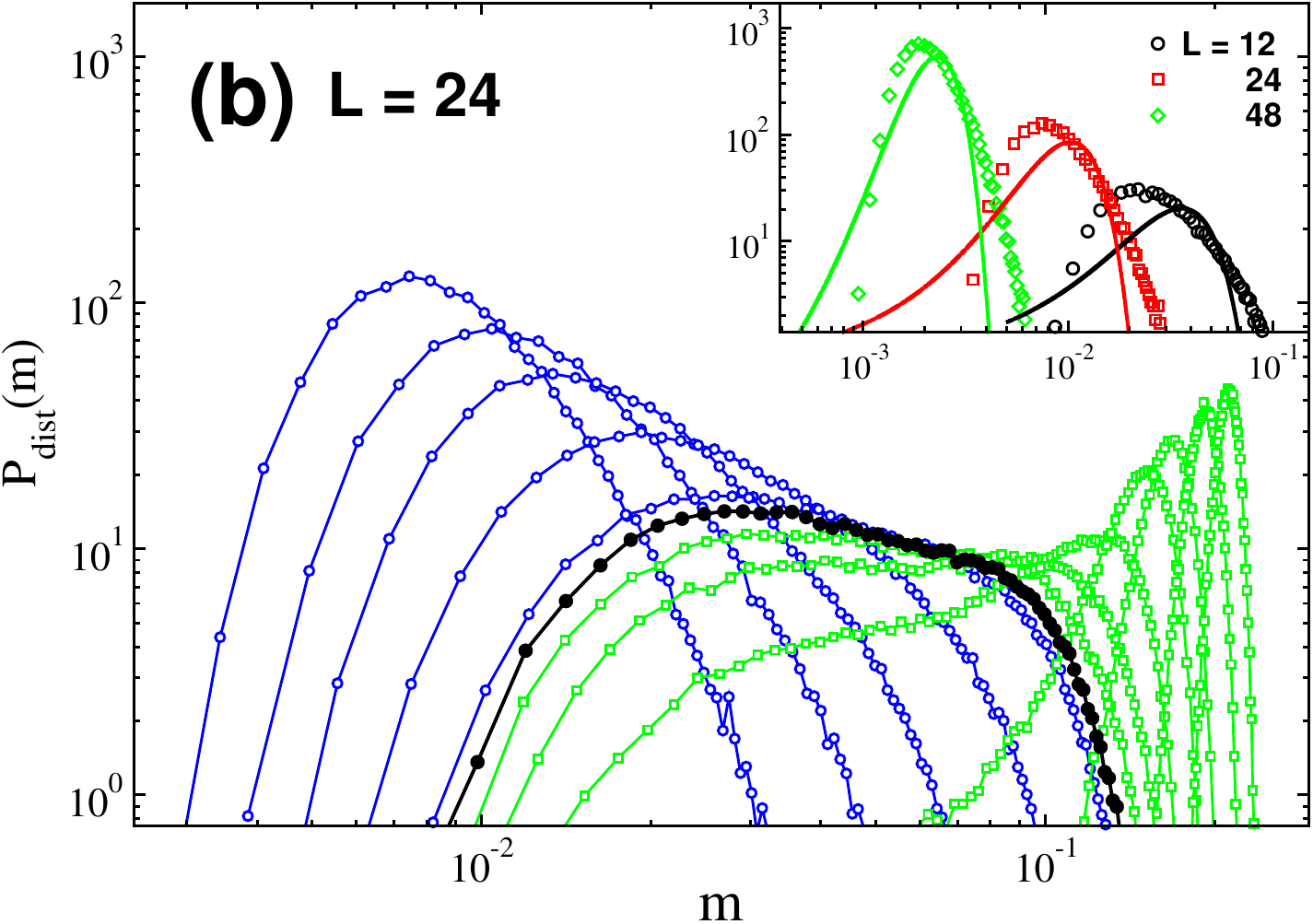}}}
	\caption{Probability distribution $P_{\text{dist}}(m)$ vs mass fraction $m$ for (a) line loops with maximal connection, and (b) FK graphs, for temperatures below (denoted by circle), above (denoted by square), and at the $T_{\rm c}$ (filled circle). The temperature increases as the maximum in $P_{\text{dist}}(m)$ shifts from left to right. The linear size of the system is fixed to $L = 24$. Insets in (a) and (b) show the probability distribution $P_{\text{dist}}(m)$ for $T = 1.27 < T_{\rm c}$ and different system sizes (see the key). The curves in solid plotted above the different datasets, represent the Gaussian distribution with mean and variance determined from those datasets.}
	\label{Sfig5c}
\end{figure}

In Fig.~\ref{Sfig5a}, the behavior of $U_{\rm 4}$ for the geometric line loops constructed with the maximal connection rule is explored. In panel (a), $U_{\rm 4}$ is plotted against temperature $T$ for different system sizes $L$. The value of $T_{\rm p}$ is obtained from the common intersection of different datasets, $T_{\rm p} \simeq 1.3135(3)$, which is in excellent agreement with $T_{\rm c}$. For $T > T_{\rm p}$, $U_{\rm 4}$ approaches a limiting value $2/3$ independently of $L$, while at low $T$ the value of $U_{\rm 4}$ decreases on lowering $L$ and even becomes negative for small $L$ and weak  $T$. This fact is in contradiction with what is expected in conventional continuous phase transitions, where $U_{\rm 4}$ approaches zero due to Gaussianity of the order parameter (kurtosis equals $3$). When magnifying the data of this panel in the inset, it is found that the different datasets intersect at a common point, which is indistinguishable from $T_{\rm c}$. This $L$-independent behavior of $U_{\rm 4}$ at $T_{\rm c}$ is a signature of continuous transition. In addition, for each $L$ a strange dip in $U_{\rm 4}$ is also observed near (but below) $T_{\rm c}$. If this dip were to increase rapidly with $L$, one should conclude that the transition is of first order. However, the inset also shows that the dip does not change a lot with the system size. In the literature~\cite{PhysRevLett.108.045702,PhysRevX.7.031052}, such a behavior was predicted due to anomalies in the order parameter distribution (see Fig.~\ref{Sfig5c}) near $T_{\rm c}$ and the transition was referred to as \textit{pseudo-first-order} transition. A more quantitation understanding is developed below.

In the panel (b) of Fig.~\ref{Sfig5a}, the scaling relation~\eqref{Seq23} is tested for the data in (a) while fixing $T_{\rm p} = T_{\rm c}$ and treating $\nu_{\rm p}$ as a free parameter. The best scaling collapse for datasets belonging to different $L$ is obtained for $1/\nu_{\rm p} \simeq 1.571(18)$. To judge the quality of the collapse, data of the main frame are magnified around the dip and common intersection in the inset, which confirms the validity of the relation~\eqref{Seq23}. Since both dip and common intersection are described by the same scaling relation~\eqref{Seq23}, it can be concluded that the observed dip is also a universal feature of the fixed point at $T_{\rm c}$. Therefore, the percolation transition in line loops is of continuous nature. A similar picture is obtained for line loops constructed from the stochastic rule. Notice that the values of $T_{\rm p}$ and $1/\nu_{\rm p}$ obtained above are in good match with the ones extracted from the spanning probability $P_{\rm S}$ in the main text.

In Fig.~\ref{Sfig5b}, the same analysis is repeated for the FK graphs. The behavior is analogous to the one of the line loops (a common intersection at $T_{\rm p} = T_{\rm c}$ and slowly varying dip right below $T_{\rm c}$ for different $L$), though the value of $U_{\rm 4}$ remains nonzero at low $T$. Furthermore, a perfect agreement is obtained for the scaling relation~\eqref{Seq23} in the whole range of $T$. This analysis gives $T_{\rm p} \simeq 1.3139(6)$ and $1/\nu_{\rm p} \simeq 1.575(15)$, which are in good agreement with those obtained from $P_{\rm S}$ in the main text.

Finally, the percolation transition in the FK graphs is also of continuous nature with characteristics of a pseudo-first-order transition.

Figure~\ref{Sfig5c} shows the probability distribution $P_{\text{dist}}(m)$ for (a) line loops and (b) FK graphs, above, below, and at $T_{\rm c}$. See the caption for details. At $T \gtrsim T_{\rm c}$ the behavior of $P_{\text{dist}}(m)$ is similar to what is expected in a conventional continuous phase transition~\cite{PhysRevE.99.042150}. However, as the temperature is lowered below $T_{\rm c}$ it shows significant differences. $P_{\text{dist}}(m)$ is not Gaussian and produces a value of kurtosis different from $3$ (see the insets). Therefore, $U_{\rm 4}$ no longer approaches zero in the disordered phase. The dip in $U_{\rm 4}$ below $T_{\rm c}$ can be related to such anomalies in $P_{\text{dist}}(m)$~\cite{PhysRevLett.108.045702,PhysRevX.7.031052}. Notice that the different behavior of $P_{\text{dist}}(m)$ at $T_{\rm c}$ in (a) and (b) is due to the localization error in $T_{\rm c}$.

\section{Fractal character of geometric objects at the critical point}
\label{ssec7}

This section explores the fractal characteristics of the different geometric objects and their substructures at the percolation threshold, which coincides with the thermodynamic critical temperature $T_{\rm c}$ (see the main text). For this purpose, we calculate the fractal dimension $D_{\rm f}$ of these objects and the backbone exponent $D_{\rm b}$ to be defined below.

\begin{figure}[t!]
	\centering
	\rotatebox{0}{\resizebox{.44\textwidth}{!}{\includegraphics{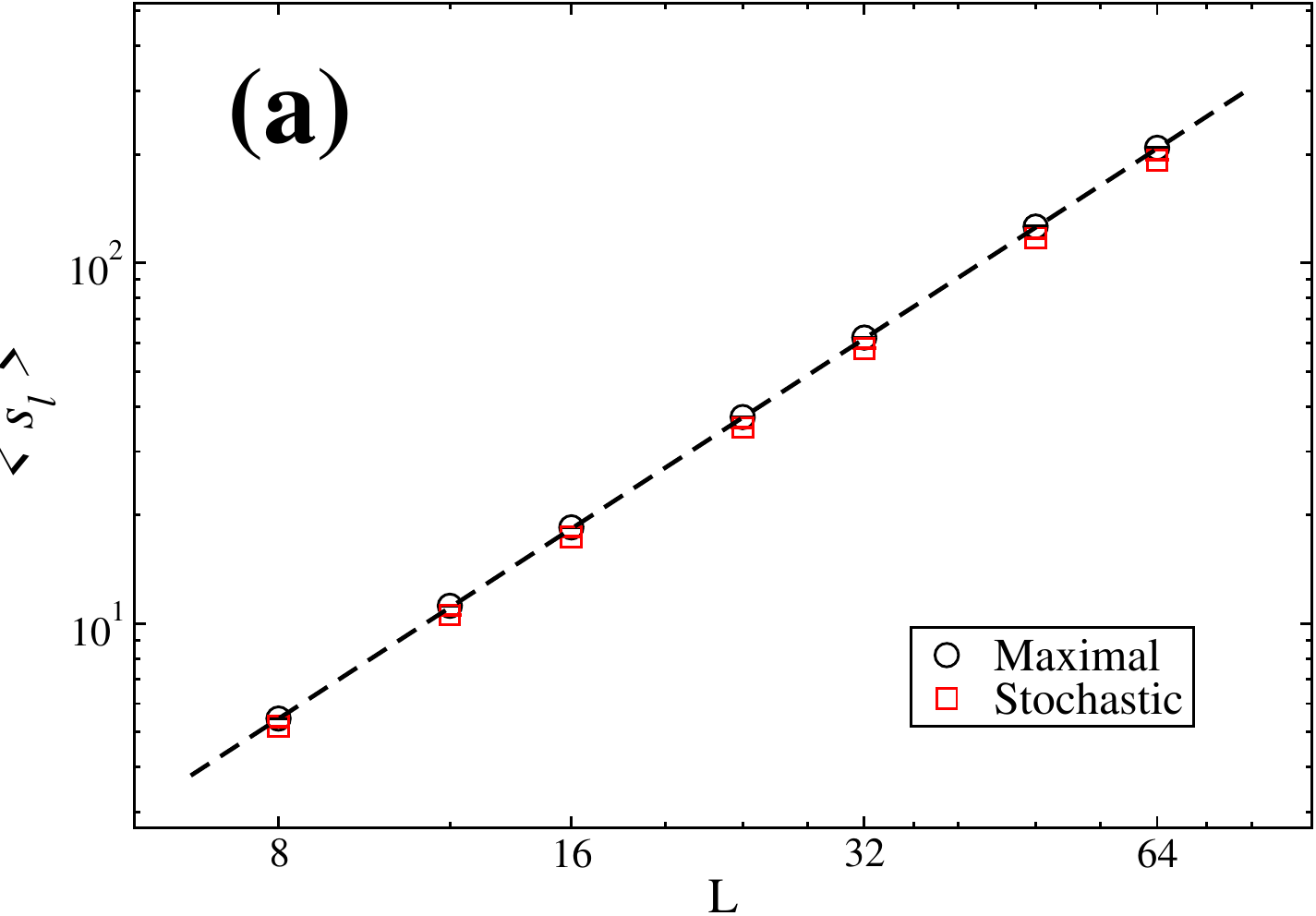}}}
	\rotatebox{0}{\resizebox{.44\textwidth}{!}{\includegraphics{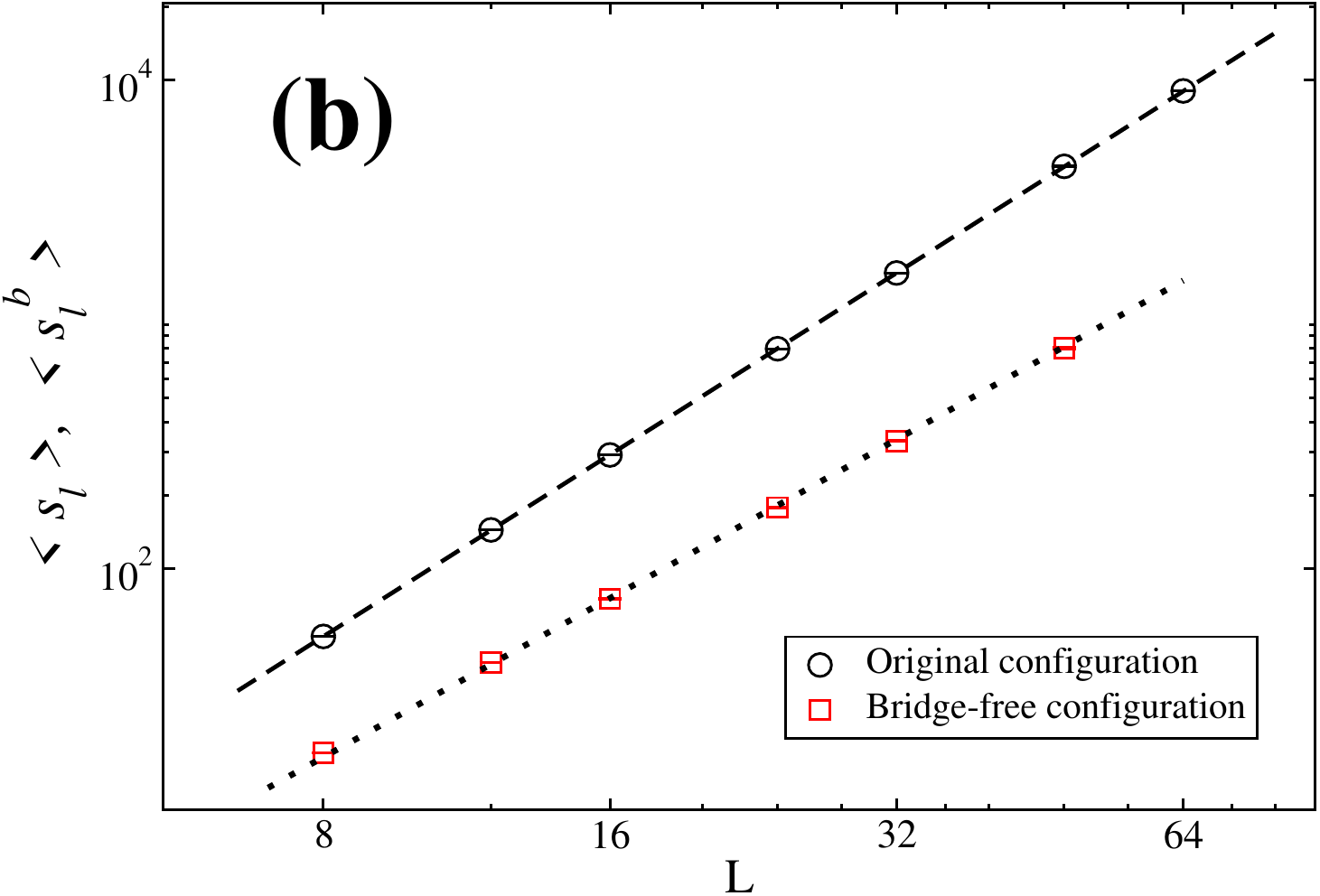}}}
	\caption{(a) Plot of $\langle s_{l} \rangle$ vs $L$ for closed line loops formed with different connection methods (see the key) at $T_{\rm c}$. The dashed line represents the best power-law fit to the law $\langle s_{l} \rangle \sim L^{D_{\rm f}}$, with $D_{\rm f} \simeq 1.74$. (b) Plot of $\langle s_{l} \rangle$ (circle) and $\langle s_{l}^{\rm b} \rangle$ (square) against $L$ for Fortuin-Kasteleyn (FK) clusters at $T_{\rm c}$. The dashed and dotted lines represent the best fit to $\langle s_{l} \rangle \sim L^{D_{\rm f}}$ and $\langle s_{l}^{\rm b} \rangle \sim L^{D_{\rm b}}$, with $D_{\rm f} = 2.48$ and $D_{\rm b} = 2.16$, respectively. The error bars in both panels are smaller than the symbol sizes.}
	\label{Sfig16}
\end{figure}

For convenience of the reader, we recall that a percolating cluster consists of many fractal substructures. In particular, a backbone is subset of those \textit{special} constituents (sites/bonds) which would carry the electric current if a voltage difference is set up across the opposite edges of the percolating cluster. The exponent $D_{\rm b}$ is associated with this backbone structure and it is useful to distinguish different fractal objects with similar value of $D_{\rm f}$ (see Refs.~\cite{PhysRevLett.53.1121,Strelniker2009} for a detailed discussion).

Here, we exploit a computationally efficient method to estimate $D_{\rm b}$. In the present model, the loops/clusters are constructed by the lines which pierce the neighboring frustrated plaquettes of the lattice. Therefore, a line passing through two neighboring plaquettes (belonging to same cube) serves as a ``bond" between them. We classify such bonds into two categories, \textit{bridge} bonds and \textit{non-bridge} bonds. A bridge bond is such that,  when removed, the cluster breaks up into two separate clusters. Otherwise, the bond is called non-bridge. In the next step, all the bridge bonds are removed from an original plaquette configuration. The configuration so obtained does not have dangling ends excluding the small to large sized blobs. The average mass $\langle s_{l}^{\rm b} \rangle$ of the largest blob gives the backbone exponent $D_{\rm b}$ as,
\be
\langle s_{l}^{\rm b} \rangle \sim L^{D_{\rm b}}
\; .
\label{Seq51}
\ee
We remind that the line loops constructed from any connection method cannot have dangling ends. Therefore, their backbone exponent $D_{\rm b}$ is equivalent to their fractal dimension $D_{\rm f}$. This is not the case for the 
lines built from the FK clusters and for them $D_{\rm b}$ can be different from $D_{\rm f}$. Moreover, $D_{\rm b}$
for the FK clusters needs not be the same as $D_{\rm f}$  for the line ones.

In Fig.~\ref{Sfig16}(a), the average mass $\langle s_{l} \rangle$ of the largest closed loops is plotted against the system size $L$ for both connection methods, maximal and stochastic. This quantity shows a power-law growth on the log-log scale of the figure, as described in Eq.~\eqref{Seq5}. We obtain $D_{\rm f} \simeq 1.738(8)$ irrespective of the connection method. This value of $D_{\rm f}$ is in excellent agreement with that of the HT expansion graphs in the 3D Ising model~\cite{PhysRevE.77.061108,shimada2016fractal,PhysRevE.101.012104} (see the discussion in the above section).

In Fig.~\ref{Sfig16}(b), $\langle s_{l} \rangle$ and $\langle s_{l}^{\rm b} \rangle$ are calculated for the FK clusters, which are extracted from the original and bridge-free configurations, respectively. From the best power-law fits to different datasets in Fig.~\ref{Sfig16}(b), we obtain $D_{\rm f} \simeq 2.481(6)$ and $D_{\rm b} \simeq 2.159(8)$, which are in agreement with the recent numerical estimates for FK clusters in 3D Ising model~\cite{PhysRevE.99.042150}.

\begin{figure}[t!]
	\centering
	\rotatebox{0}{\resizebox{.44\textwidth}{!}{\includegraphics{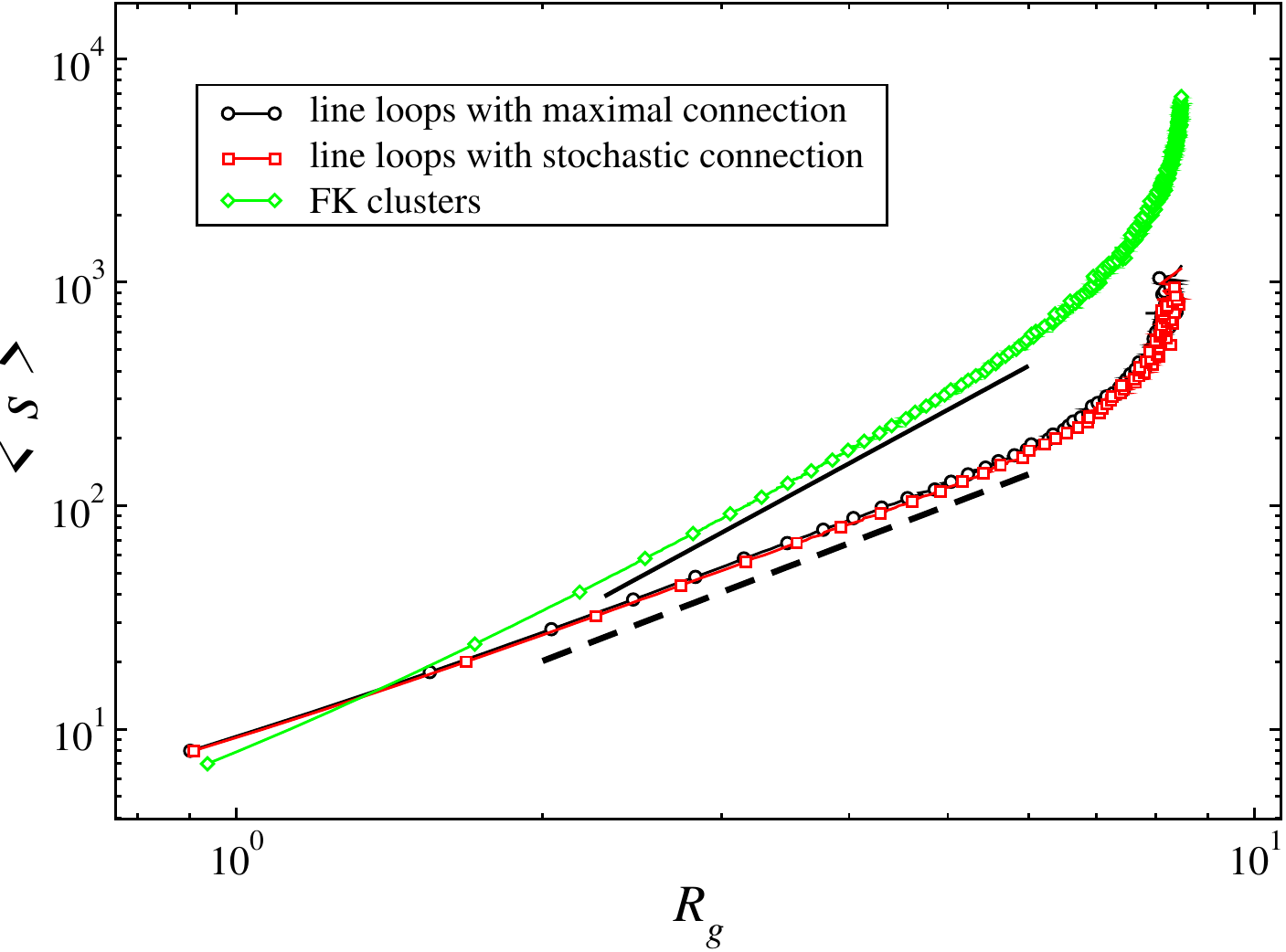}}}
	\caption{Average mass $\langle s \rangle$ against the radius of gyration $R_{\rm g}$, for a pure gauge system of linear size $L=24$ at $T = T_{\rm c}$. Different datasets represent different geometric objects (see the key). The solid and dashed lines denote the law $\langle s \rangle \sim R_{\rm g}^{D_{\rm f}}$, with fractal dimensions $D_{\rm f} = 1.74$ and $D_{\rm f} = 2.48$, respectively.}
	\label{Sfig18}
\end{figure}

For a consistency check, we also calculate the fractal dimension $D_{\rm f}$ from the standard box counting approach, i.e., using Eq.~\eqref{Seq6}. We plot the average mass $\langle s \rangle$ of the different geometric objects (represented by different datasets) against their radius of gyration $R_{\rm g}$ in Fig.~\ref{Sfig18}, for a system of linear size $L=24$. On the scales larger than the lattice spacing ($R_{\rm g} > 1$), a quite good agreement is observed with the above estimates of $D_{\rm f}$. Notice that the bend in datasets at large $R_{\rm g}$ is due to finite size effects and goes away with increase in system size.

\section{Statistics of line loops at different temperatures}
\label{ssec8}

Equation~\eqref{Seq3} reveals that at the percolation temperature $T_{\rm p}$ (which coincides with the 
critical temperature $T_{\rm c}$ for the line loops; see the main text) the number density $N(s)$ of loops with length 
$s$ falls in a power law fashion, $N(s)/L^3 \propto s^{-\tau_{\rm p}}$, as the tension vanishes, $\epsilon =0$, at  $T = T_{\rm p}$. This can be expressed in terms of a finite size scaling function as,
\be
\label{Seq11}
N(s)/L^3 = s^{-\tau_{\rm p}} G\left( s/L^{D_{\rm f}} \right)
\; ,
\ee
where $D_{\rm f}$ is the fractal dimension. In the main text and Sec.~\ref{ssec5} the above laws are 
justified for maximal and stochastic connection methods of the loops, respectively. The law $N(s)/L^3 \propto s^{-\tau_{\rm p}}$ is also confirmed with exponent $\tau_{\rm p} \simeq 2.715(15)$ for the line loops, irrespective of the connection rule used.

\begin{figure}[t!]
	\centering
	\rotatebox{0}{\resizebox{.48\textwidth}{!}{\includegraphics{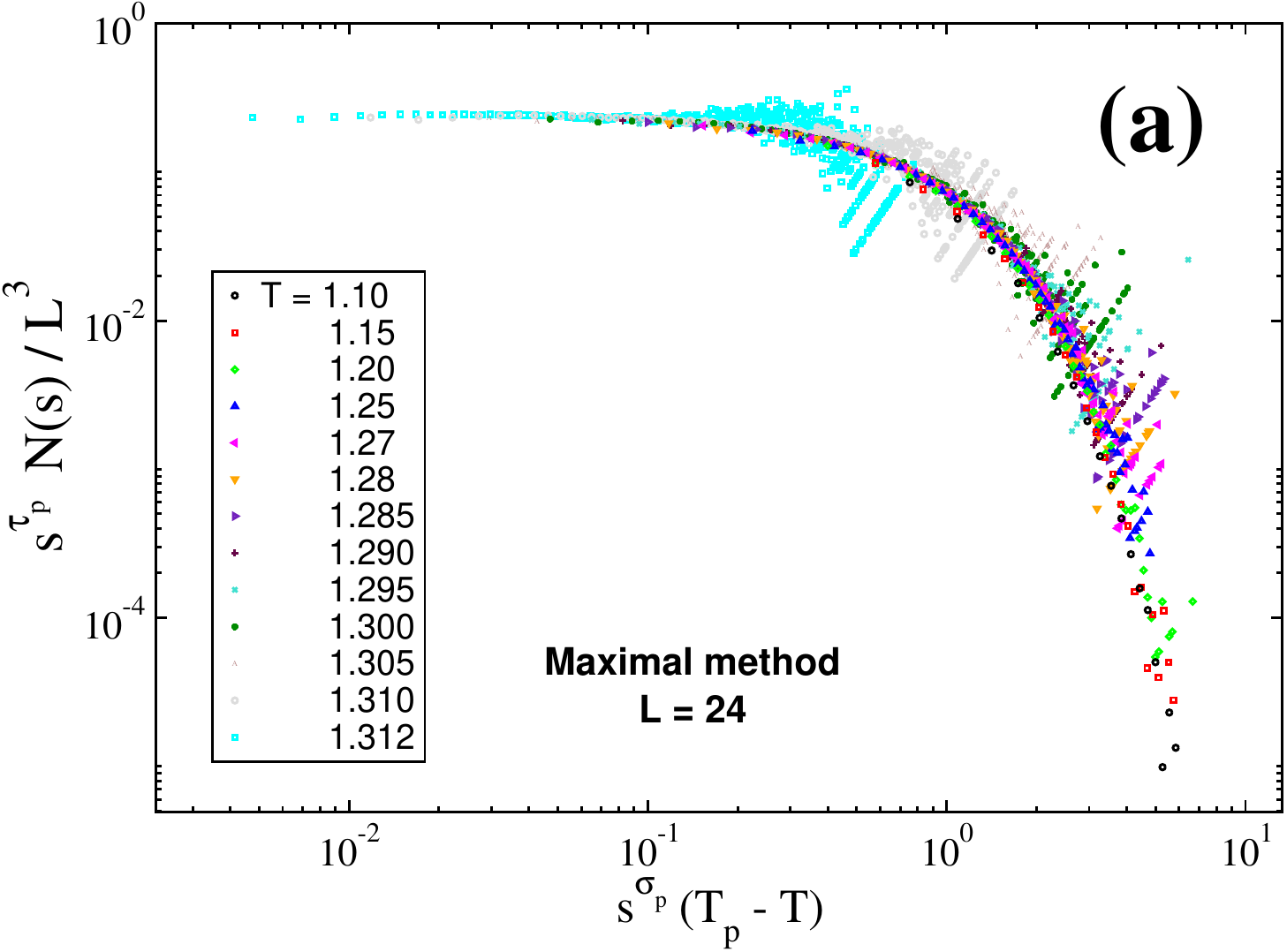}}}
	\rotatebox{0}{\resizebox{.48\textwidth}{!}{\includegraphics{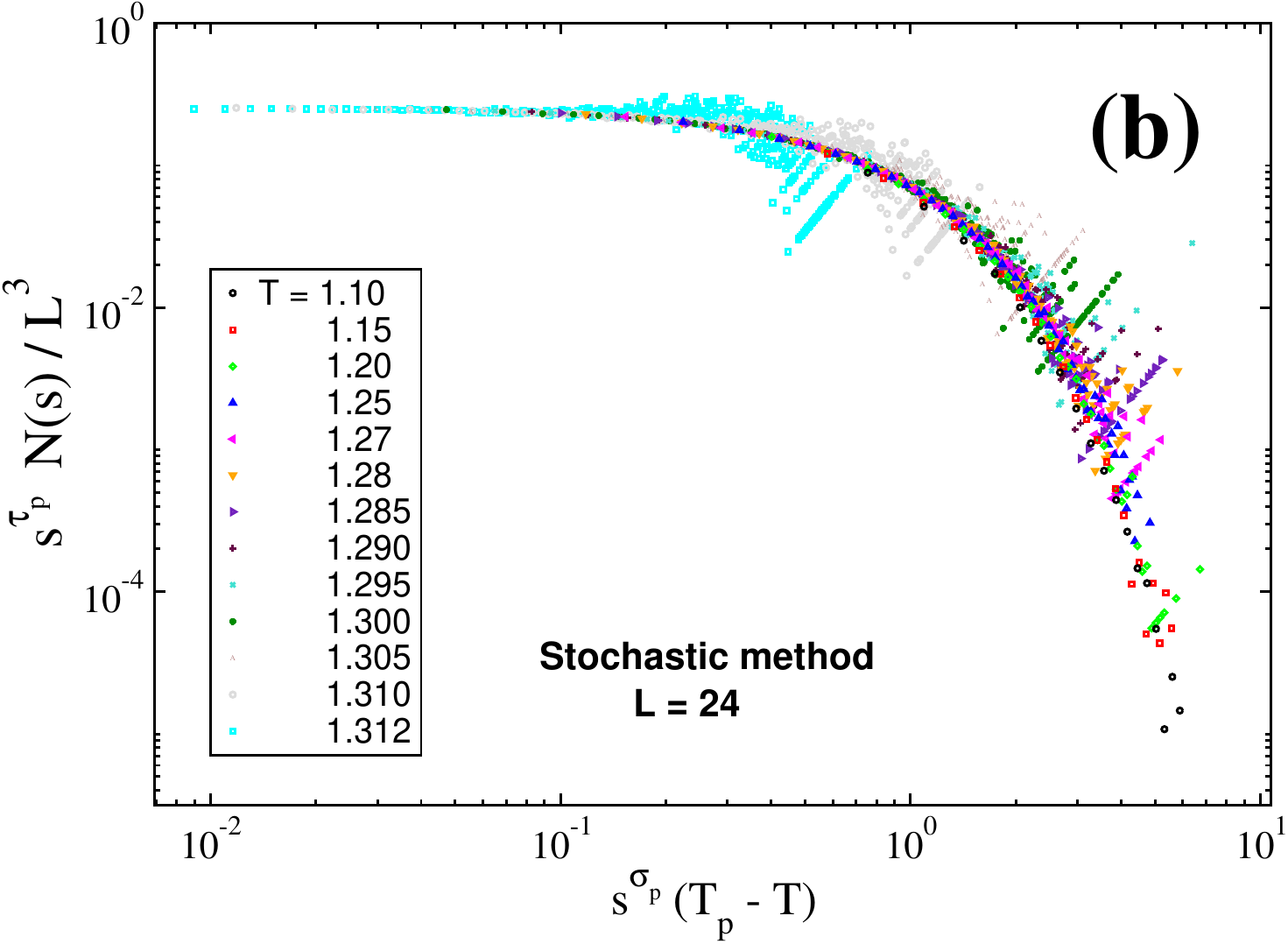}}}
	\caption{Scaling plots of $s^{\tau_{\rm p}} N(s)/L^3$ vs $s^{\sigma_{\rm p}}(T_{\rm p}-T)$ for a system of linear size $L=24$ at different temperatures below $T_{\rm c}$ (see the key): (a) maximal and (b) stochastic connection of closed line loops. Herein, $\tau_{\rm p}$ is fixed to $\tau_{\rm p} = 2.71$.}
	\label{Sfig6}
\end{figure}

We recall that for temperatures below $T_{\rm c}$, Eq.~\eqref{Seq3} remains valid but with a nonzero tension $\epsilon$. We reformulate the scaling law~\eqref{Seq3} as,
\be
N(s)/L^3 \simeq s^{-\tau_{\rm p}} f_{\rm N}\left( s^{\sigma_{\rm p}} \vert T - T_{\rm p} \vert \right)
\; ,
\label{Seq12}
\ee
where $f_{\rm N}$ is a scaling function. The above relation allows us to extract the value of the exponent $\sigma_{\rm p}$ and cross check with the obtained value of the exponent $\nu_{\rm p}$ in the main text, as $\sigma_{\rm p} = 1/(D_{\rm f} \nu_{\rm p})$. In Figs.~\ref{Sfig6}(a) and (b), the above law is checked for maximal and stochastic connections, respectively, where the size of the system is $L = 24$ and different datasets correspond to temperatures below $T_{\rm c}$. Here, the value of $\tau_{\rm p}$ is fixed to $\tau_{\rm p} \simeq 2.71$ (as obtained above), and the value of the exponent $\sigma_{\rm p}$ is chosen so as to enable a good scaling collapse for different datasets. We obtain
\begin{equation}
	\sigma_{\rm p} \simeq 0.91(1)
\end{equation}
in both figures. (The large $s$ deviation from the collapse is a finite size effect, as the scaling law~\eqref{Seq12} is valid for the small loops only.)

\begin{figure}[t!]
	\centering
	\rotatebox{0}{\resizebox{.48\textwidth}{!}{\includegraphics{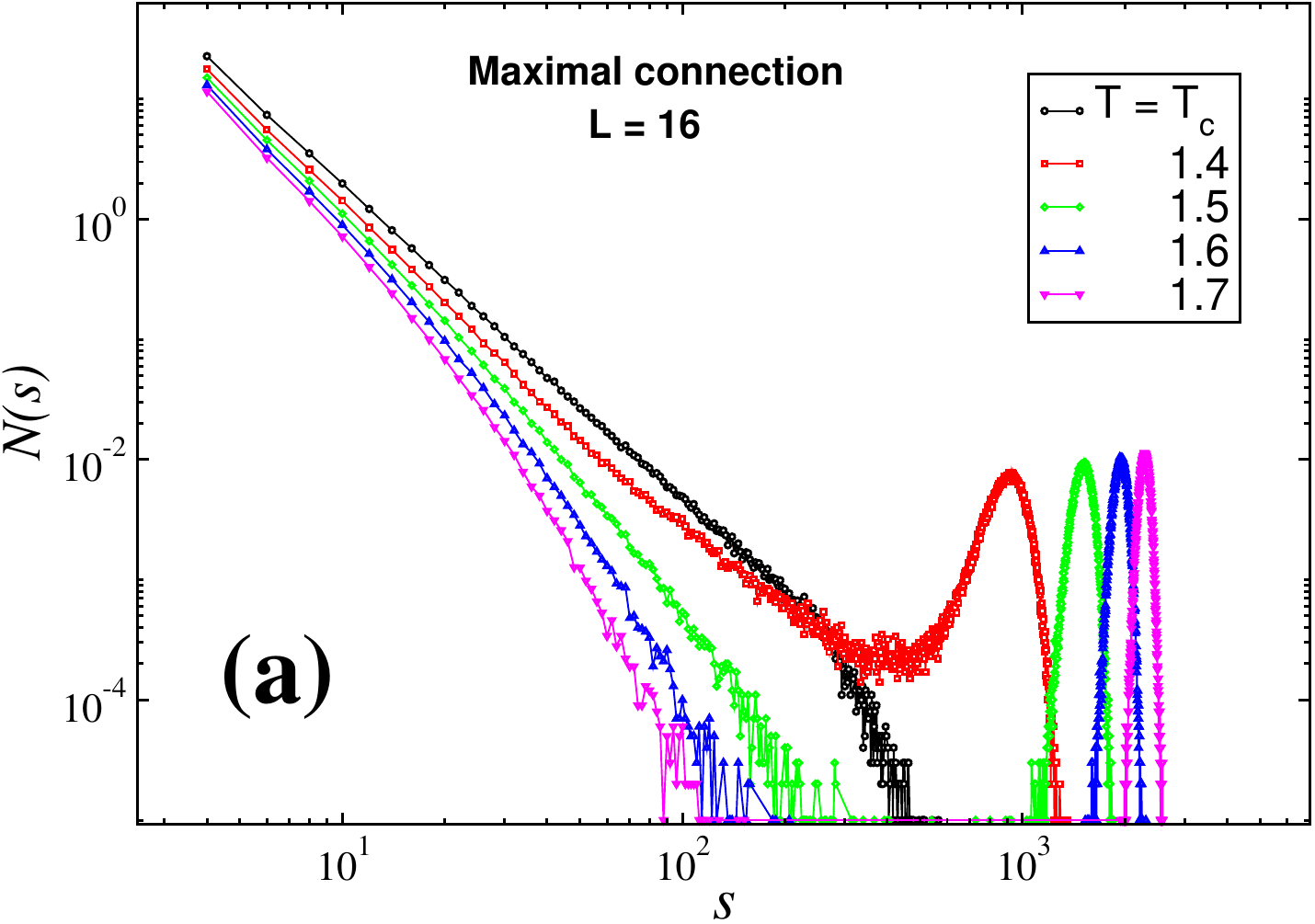}}}
	\rotatebox{0}{\resizebox{.48\textwidth}{!}{\includegraphics{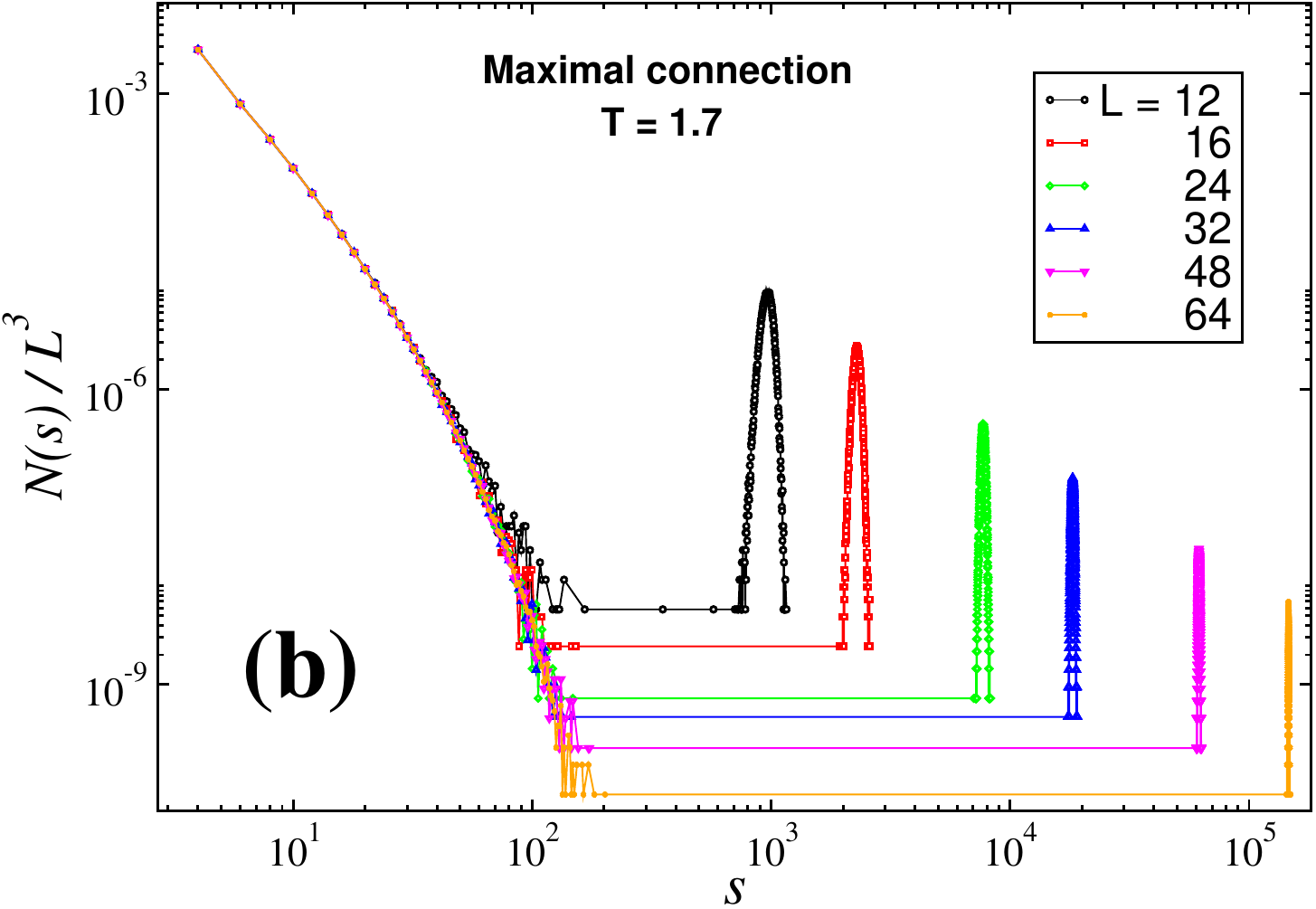}}}
	\caption{Closed loops with maximal connection. (a) Plot of $N(s)$ vs $s$ for a system of linear size $L=16$ at different temperatures above and at $T_{\rm c}$ (see the key). (b) Plot of $N(s)/L^3$ vs $s$ for different system sizes $L$ (see the key) and temperature $T=1.7$.}
	\label{Sfig7}
\end{figure}

\begin{figure}[h!]
	\centering
	\rotatebox{0}{\resizebox{.48\textwidth}{!}{\includegraphics{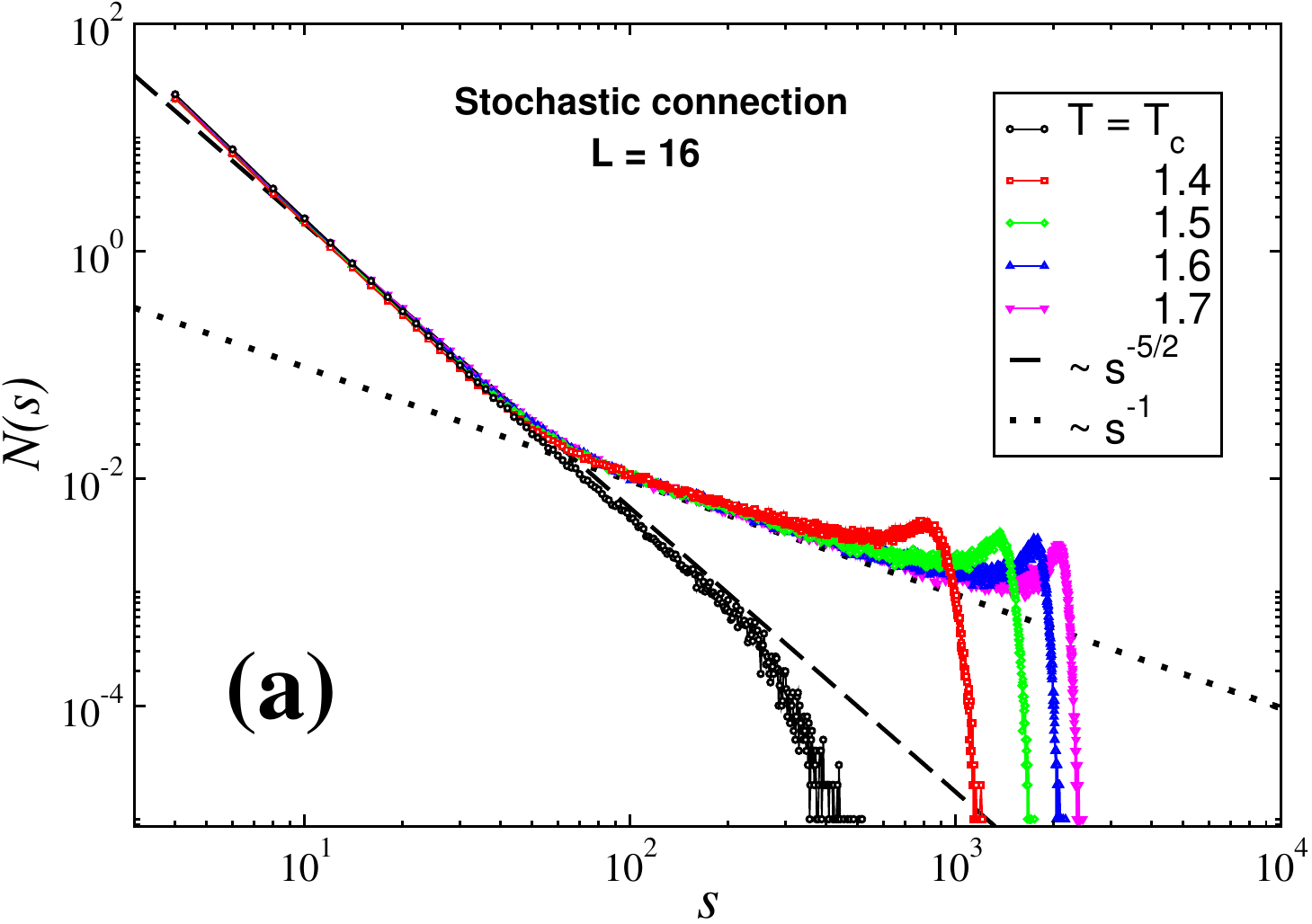}}}
	\rotatebox{0}{\resizebox{.48\textwidth}{!}{\includegraphics{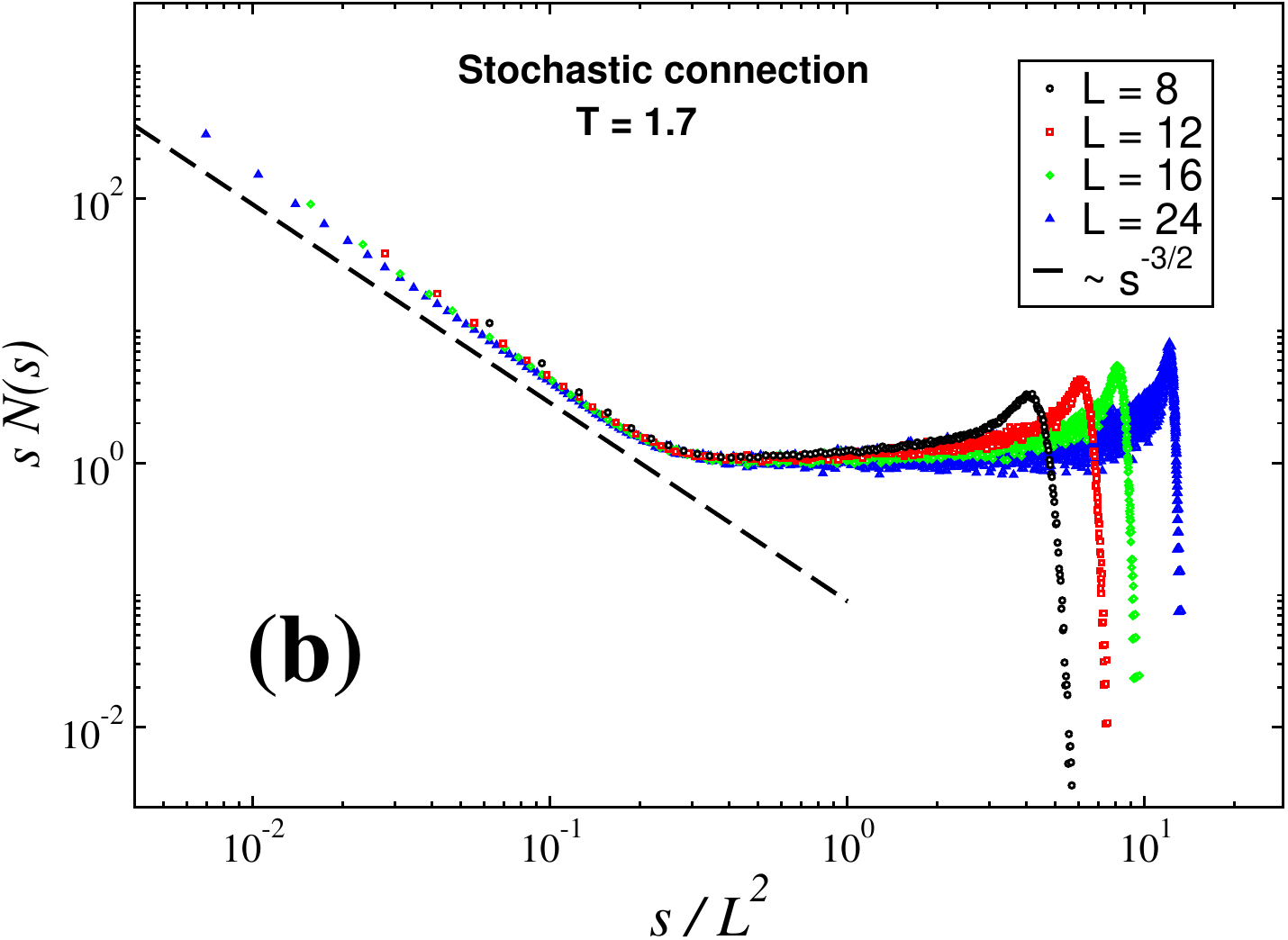}}}
	\caption{Closed loops with stochastic connection. (a) Plot of $N(s)$ vs $s$ for a system of linear size $L=16$ at different temperatures above and at $T_{\rm c}$ (see the key). The dashed and dotted lines represent the laws $N(s) \sim s^{-5/2}$ and $N(s) \sim s^{-1}$, respectively. (b) Plot of $s N(s)$ vs $s/L^2$ for different system sizes $L$ (see the key) and temperature $T=1.7$. The dashed line represents the power law $y(x) \sim x^{-3/2}$.}
	\label{Sfig8}
\end{figure}

Now we finally discuss the statistics of line loops at high temperatures ($T \gg T_{\rm c}$), for which some generic predictions for $N(s)$ are available from the so-called fully-packed loop models. The equilibrium configuration at high temperatures is filled with system spanning line defects. If the configuration is so packed that two line defects enter and exit each elementary cube, the number density $N(s)$ of geometrical loops of length $s$ behaves as~\cite{PhysRevLett.107.177202,PhysRevLett.111.100601},
\bea
N(s)/L^3 &\simeq& 
\left\{
\begin{array}{l}
	s^{-5/2}
	\; ,
	~~~~~~~~ s \ll L^2
	\; ,
	\\
	s^{-1} L^{-3}
	\; ,
	~~~~~ s \gg L^2
	\; .
\end{array}
\right.
\label{Seq4}
\eea
The small scale regime $s \ll L^2$ corresponds to the result for a Gaussian random walk~\cite{de1979scaling}, while the large scale regime supports the fully-packed loop model prediction~\cite{PhysRevLett.107.177202,PhysRevLett.111.100601}.

\begin{figure}[h!]
	\centering
	\rotatebox{0}{\resizebox{.44\textwidth}{!}{\includegraphics{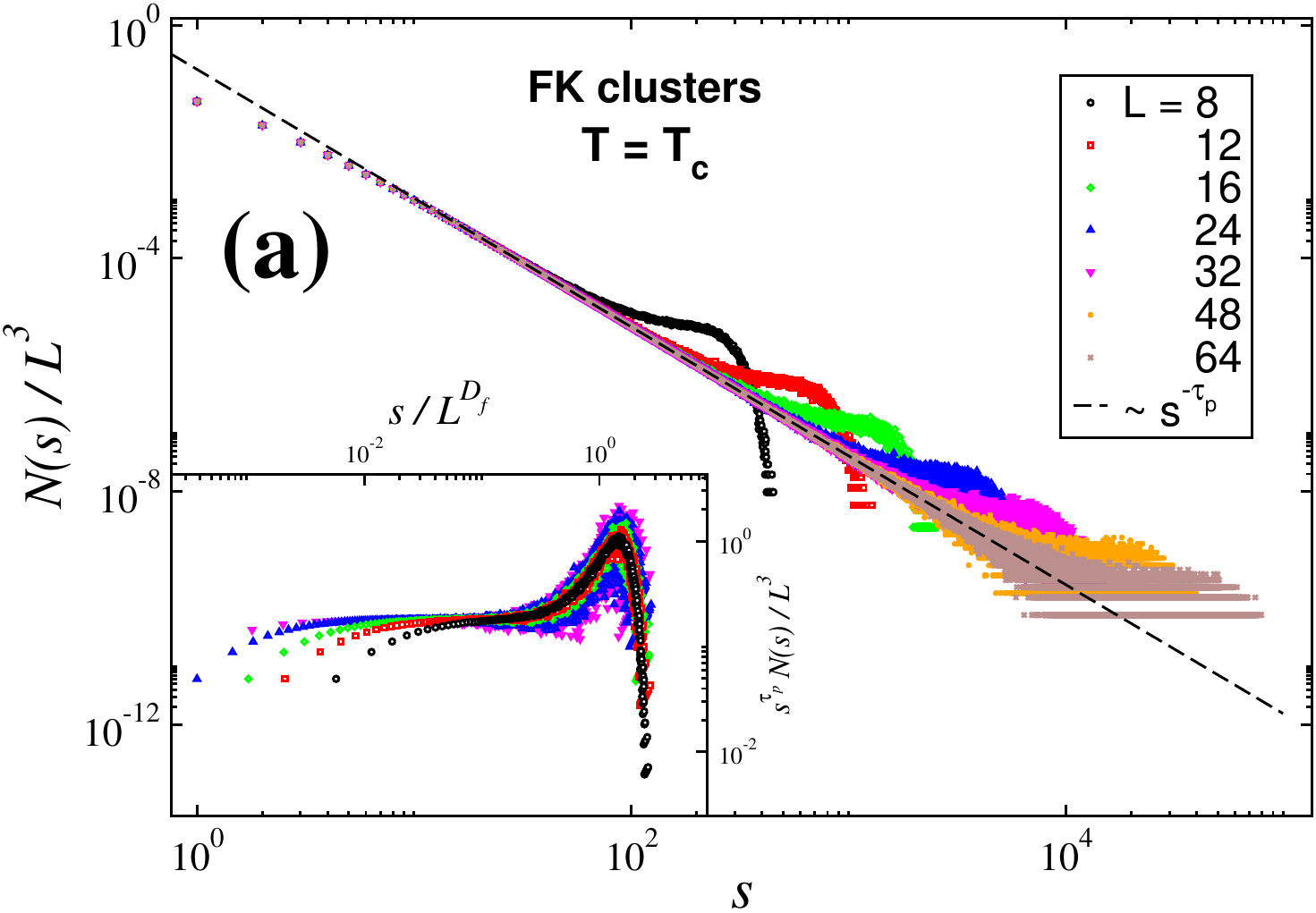}}}
	\rotatebox{0}{\resizebox{.44\textwidth}{!}{\includegraphics{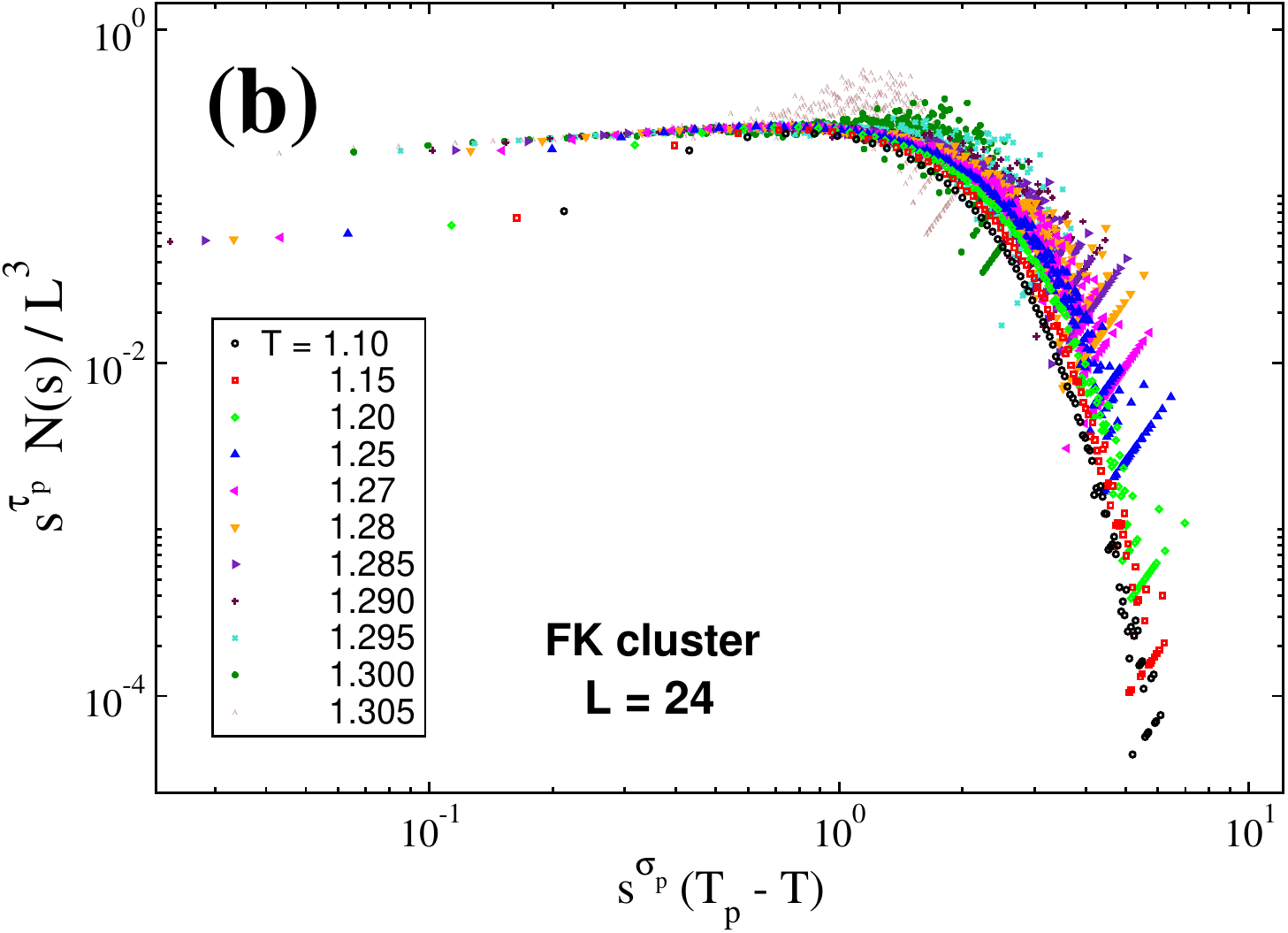}}}
	\caption{FK clusters: (a) Plot of $N(s)/L^3$ vs $s$ for different system sizes $L$ (see the key) at the critical temperature $T_{\rm c}$. The dashed line represents the law $N(s)/L^3 \propto s^{-\tau_{\rm p}}$, with $\tau_{\rm p} = 2.21$. In the inset, the scaling variable $s^{\tau_{\rm p}} N(s)/L^3$ is plotted against $s/L^{D_{\rm f}}$ for the data in main frame, with $\tau_{\rm p}$ and $D_{\rm f}$ fixed to $\tau_{\rm p} = 2.21$ and $D_{\rm f} = 2.48$. (b) Scaling plot of $s^{\tau_{\rm p}} N(s)/L^3$ vs $s^{\sigma_{\rm p}}(T_{\rm p}-T)$ for a system of linear size $L=24$ at different temperatures below $T_{\rm c}$ (see the key), with $\tau_{\rm p} = 2.21$.}
	\label{Sfig19}
\end{figure}

Here, we test the predictions in Eq.~\eqref{Seq4} for the line loops in the present model  formed by different connection methods. In Fig.~\ref{Sfig7}(a), the quantity $N(s)$ is plotted against $s$ for the maximal connection rule, for a system with linear
size $L = 16$. In contrast to the prediction in Eq.~\eqref{Seq4}, $N(s)$ indicates an exponential fall at small $s$, while only a peak is obtained at large $s$. The position of the peak increases with the temperature $T$ above $T_{\rm c}$. In Fig.~\ref{Sfig7}(b), the behavior of the quantity $N(s)$ is tested for different system sizes while fixing temperature to a higher $T = 1.7$. With increase in $L$, the regime of initial exponent fall increases. Apart from that, the peak position is also shifted to large $s$. This actually demonstrates that due to the maximal connection criterion, the system at high temperatures is evaded by a very large loop (which wraps around the periodic boundaries several times). Other remaining loops are very small (both in size and number) and therefore, they cannot produce the law~\eqref{Seq4}.

Instead, the predictions in Eq.~\eqref{Seq4} are nicely verified by the line loops constructed with the stochastic method, as shown in Fig.~\ref{Sfig8}. In Fig.~\ref{Sfig8}(a), $N(s)$ is plotted against $s$ for $L = 16$. The data in this figure clearly show agreement with the law~\eqref{Seq4}: $N(s) \sim s^{-5/2}$ for small $s$ and $N(s) \sim s^{-1}$ for large $s$. The same behavior is also found for other system sizes (not shown here for brevity). To confirm the two power laws in Eq.~\eqref{Seq4} more explicitly, the quantity $s N(s)$ is plotted against $s/L^2$ for different system sizes and $T = 1.7$ in Fig.~\ref{Sfig8}(b). A power law decay $s^{-3/2}$ at $s \ll L^2$ and a constant behavior at $s \gg L^2$ is observed in such kind of plot, which confirms the validity of \eqref{Seq4}.

The results in this section are quantitatively similar to the ones in~Refs.~\cite{Kobayashi_2016,PhysRevE.94.062146}.

\section{Number density of FK clusters}
\label{ssec10}

To extract the critical exponents $\tau_{\rm p}$ and $\sigma_{\rm p}$ for the FK clusters, one can borrow the analysis presented above for the line loops, i.e., to exploit the finite-size scaling function~\eqref{Seq11} and law~\eqref{Seq12} for the number density $N(s)$. We remind that similarly to the line loops, the percolation temperature $T_{\rm p}$ of the FK clusters also coincides with the critical temperature $T_{\rm c}$; see the main text.

In the main frame of Fig.~\ref{Sfig19}(a), the normalized number density $N(s)/L^3$ of the FK clusters is plotted against their mass $s$ for different system sizes $L$ at the critical temperature $T = T_{\rm c}$. As expected from Eq.~\eqref{Seq3}, this quantity decays algebraically at $T_{\rm c}$, $N(s)/L^3 \sim s^{-\tau_{\rm p}}$, with
\begin{equation}
	\tau_{\rm p} \simeq 2.207(5)
	\; .
\end{equation}
This value of $\tau_{\rm p}$ gives $D_{\rm f} \simeq 2.48$, which is in agreement with our direct probe of $D_{\rm f} [\simeq 2.481(6)]$ from $\langle s_{l} \rangle$. In the inset of Fig.~\ref{Sfig19}(a), the validity of the function~\eqref{Seq11} is tested when fixing $\tau_{\rm p} \simeq 2.21$ and $D_{\rm f} \simeq 2.48$. The deviation from the scaling collapse at small $s$ (on the order of the lattice spacing) is obvious as the function~\eqref{Seq11} is valid only for the fractal clusters.

In Fig.~\ref{Sfig19}(b), the scaling law~\eqref{Seq12} is tested by plotting the quantity $s^{\tau_{\rm p}} N(s)/L^3$ against the variable $s^{\sigma_{\rm p}} (T_{\rm p} - T)$ for the FK clusters in a system of linear size $L = 24$. We fixed $\tau_{\rm p}$ to $\tau_{\rm p} \simeq 2.21$, and the exponent $\sigma_{\rm p}$ is estimated by enabling the good scaling collapse of the different datasets. Notice that the function $f_{\rm N}$ in Eq.~\eqref{Seq12} accounts for how the number density $N(s)$ of cluster size $s$ changes near $T_{\rm c}$, but it does not consider the role of the system size $L$. Therefore, finite size corrections are expected in this law. We find
\begin{equation}
	\sigma_{\rm p} \simeq 0.64(1)
	\; ,
\end{equation}
which is also in agreement with the definition $\sigma_{\rm p} = 1/ (D_{\rm f} \nu_{\rm p})$.

\end{document}